\documentclass{vldb}
\usepackage{graphicx}
\usepackage{hyperref}
\usepackage{balance}  
\usepackage{xspace}
\usepackage{bm} 

\usepackage{tikz}
\usetikzlibrary{fit,shapes, arrows, shadows, calc, automata, topaths, positioning,matrix,backgrounds,decorations.pathmorphing,backgrounds,petri,decorations.markings,decorations.pathreplacing}
\usepackage{pgfplots}
\usepackage{tcolorbox}
\usepackage{colortbl}
\usepackage{booktabs,quoting}

\usetikzlibrary{shapes.callouts}
\tikzset{
  notice/.style  = { draw, rounded corners=.1cm, rectangle callout, callout relative pointer={#1} },
}

\tikzstyle{path} = [->,double,rounded corners=.1cm]
\tikzstyle{data} = [draw, rectangle, rounded corners = .07cm, align=center, inner sep = .2cm, outer sep = .1 cm]
\tikzstyle{processor} = [draw, ellipse, align=center, inner sep = .1cm, outer sep = .1 cm]
\colorlet{clr_outofDB}{red!80!black}
\colorlet{clr_inDB}{green!50!black}
\colorlet{clr_inDB_FD}{blue}

\newcommand{\drawtable}[5]
{
   \begin{scope}[shift={(-#1/2,-#2/2)}]
      \pgfmathsetmacro{\cellwidth}{#1/#3}
      \pgfmathsetmacro{\cellheight}{#2/#4}
      \draw[fill=#5!30] (0, #2) rectangle (#1, #2-\cellheight);
      \foreach \i in {1,...,#3}
         \draw[thin, #5!70] (\cellwidth*\i, 0) -- (\cellwidth*\i, #2);
      \foreach \i in {1,...,#4}
         \draw[thin, #5!70] (0, \cellheight*\i) -- (#1, \cellheight*\i);
      \draw[#5] (0, #2-\cellheight) -- (#1, #2-\cellheight);
      \draw[thick, #5] (0, 0) rectangle (#1, #2);
   \end{scope}
}

\newcommand{\drawcylinder}[4]
{
   \begin{scope}[shift={(0,#3/2)}]
      \draw[#4] (0, #2/2) ellipse (#1/2 and #3);
      \draw[#4] (#1/2, -#2/2) arc (0:-180:#1/2 and #3);
      \draw[#4] (-#1/2, #2/2) -- (-#1/2, -#2/2);
      \draw[#4] (#1/2, #2/2) -- (#1/2, -#2/2);
   \end{scope}
}

\newcommand{\drawfilter}[2]
{
   \begin{scope}[scale=#1/7, shift={(-3.5,3)}]
      \draw[#2] (0,0)--(7, 0)--(4, -3)--(4,-5)--(3,-6)--(3,-3)--cycle;
   \end{scope}
}

\definecolor{Purple}{HTML}{911146}
\definecolor{Orange}{HTML}{CF4A30}
\definecolor{Bittersweet}{HTML}{C87A2F}
\definecolor{mMediumBrown}{HTML}{C87A2F}   

\definecolor{TolDarkPurple}{HTML}{332288}
\definecolor{TolDarkBlue}{HTML}{6699CC}
\definecolor{TolLightBlue}{HTML}{88CCEE}
\definecolor{TolLightGreen}{HTML}{44AA99}
\definecolor{TolDarkGreen}{HTML}{117733}
\definecolor{TolDarkBrown}{HTML}{999933}
\definecolor{TolLightBrown}{HTML}{DDCC77}
\definecolor{TolDarkRed}{HTML}{661100}
\definecolor{TolLightRed}{HTML}{CC6677}
\definecolor{TolLightPink}{HTML}{AA4466}
\definecolor{TolDarkPink}{HTML}{882255}
\definecolor{TolLightPurple}{HTML}{AA4499}

\definecolor{airforceblue}{rgb}{0.36, 0.54, 0.66}
\definecolor{oxfordblue}{rgb}{0, 0.33, 0.71}
\definecolor{amethyst}{rgb}{0.6, 0.4, 0.8}
\definecolor{amaranth}{rgb}{0.9, 0.17, 0.31}
\definecolor{amber}{rgb}{1.0, 0.49, 0.0}
\definecolor{applegreen}{rgb}{0.55, 0.71, 0.0}

\definecolor{darkorange}{rgb}{1.0, 0.55, 0.0}
\definecolor{forestgreen}{rgb}{0.13, 0.55, 0.13}
\definecolor{maroon}{rgb}{0.5, 0.0, 0.0}

\definecolor{darkorange}{rgb}{1.0, 0.55, 0.0}
\definecolor{forestgreen}{rgb}{0.13, 0.55, 0.13}
\definecolor{maroon}{rgb}{0.5, 0.0, 0.0}

\definecolor{lightblue}{rgb}{0.55,0.72,0.97}
\definecolor{lightyellow}{rgb}{1,1,0.4} 
\definecolor{lightorange}{rgb}{1,0.7,0.5}
\definecolor{light}{gray}{0.85}
\definecolor{heavy}{gray}{0.35}
\definecolor{goodgreen}{rgb}{0.1, 0.5, 0.1}
\definecolor{dgreen}{rgb}{0.1, 0.5, 0.1}

\definecolor{lightred}{rgb}{1,0.7,0.7}
\definecolor{lightgreen}{rgb}{0.7,1,0.7}

\definecolor{burntorange}{rgb}{0.8, 0.33, 0.0}
\definecolor{lbgreen}{RGB}{96,119,117}
\definecolor{nicepurple}{rgb}{0.76, 0.14, 0.76}

\colorlet{dred}{red!80!black}
\colorlet{dgreen}{green!50!black}
\colorlet{dorange}{orange!90!black}
\colorlet{dblue}{blue!80!black}
\colorlet{dpurple}{nicepurple!80!black}
\colorlet{dmagenta}{magenta!80!black}

\colorlet{hl_color}{dred}
\colorlet{sch_color}{oxfordblue}
\colorlet{agg_color}{goodgreen}
\colorlet{trie_color}{dorange}
\colorlet{dl_color}{dpurple}

\newcommand{\maketableo}[8]
{
  \begin{scope}[shift={(#6,#7)}, scale=1, every node/.style={transform shape}]
  \begin{scope}[shift={(-#1/2,-#2/2)}]
    \pgfmathsetmacro{\cellwidth}{#1/#3}
    \pgfmathsetmacro{\cellheight}{#2/#4}
    \draw[thick, #5, fill=white] (0, 0) rectangle (#1, #2);
    \draw[fill= #8!30] (0, #2) rectangle (#1, #2-\cellheight);
    \foreach \i in {1,...,#3}
    \draw[thin,  #5!70, fill=white] (\cellwidth*\i, 0) -- (\cellwidth*\i, #2);
    \foreach \i in {1,...,#4}
    \draw[thin, fill=white, #5!70] (0, \cellheight*\i) -- (#1, \cellheight*\i);
    \draw[#5, fill=white] (0, #2-\cellheight) -- (#1, #2-\cellheight);
  \end{scope}
  \end{scope}
}

\newcommand{\maketable}[7]
{
  \begin{scope}[shift={(#6,#7)}, scale=1, every node/.style={transform shape}]
  \begin{scope}[shift={(-#1/2,-#2/2)}]
    \pgfmathsetmacro{\cellwidth}{#1/#3}
    \pgfmathsetmacro{\cellheight}{#2/#4}
    \draw[fill=dgreen!30] (0, #2) rectangle (#1, #2-\cellheight);
    \foreach \i in {1,...,#3}
    \draw[thin, #5!70] (\cellwidth*\i, 0) -- (\cellwidth*\i, #2);
    \foreach \i in {1,...,#4}
    \draw[thin, #5!70] (0, \cellheight*\i) -- (#1, \cellheight*\i);
    \draw[#5] (0, #2-\cellheight) -- (#1, #2-\cellheight);
    \draw[thick, #5] (0, 0) rectangle (#1, #2);
  \end{scope}
  \end{scope}
}

\vldbTitle{The Relational Data Borg is Learning}
\vldbAuthors{Dan Olteanu}
\vldbDOI{https://doi.org/10.14778/3415478.3415572}
\vldbVolume{13}
\vldbNumber{12}
\vldbYear{2020}

\newcommand{\nop}[1]{}

\newcommand{\mv}[1]{\mathbf{#1}}
 
\newcommand{\RING}{\textnormal{\bf D}\xspace}
\newcommand{\RINGPLUS}{+}
\newcommand{\RINGPROD}{*}
\newcommand{\RINGZERO}{\bm{0}}
\newcommand{\RINGONE}{\bm{1}}

\newcommand{\code}[1]{\texttt{#1}}
\newcommand{\whilecond}{\code{ not converged }}

\newcommand{\dom}[1]{\code{sup(}#1\code{)}}

\newcommand{\col}[1]{\textbf{#1}}
\newcommand{\colm}[1]{\bm #1}

\newcommand{\sspace}{\text{ }}
\newcommand{\lett}{\code{let }}
\newcommand{\inn}{\code{in }}
\newcommand{\letbinding}[3]{\lett #1 = #2\sspace\inn #3}

\makeatletter
\DeclareRobustCommand\bigop[1]{%
  \mathop{\vphantom{\sum}\mathpalette\bigop@{#1}}\slimits@
}
\newcommand{\bigop@}[2]{%
  \vcenter{%
    \sbox\z@{$#1\sum$}%
    \hbox{\resizebox{\ifx#1\displaystyle.9\fi\dimexpr\ht\z@+\dp\z@}{!}{$\m@th#2$}}%
  }%
}
\makeatother

\newcommand{\biglam}{\DOTSB\bigop{\lambda}}
\newcommand{\bigsum}{\DOTSB\bigop{\Sigma}}
\newcommand{\dictbuild}[2]{\text{$\biglam\limits_{#1}#2$}}
\newcommand{\summation}[2]{\text{$\bigsum\limits_{#1}#2$}}

\begin{document}

\pagenumbering{gobble}



\title{The Relational Data Borg is Learning}


%
%
%
%

\numberofauthors{1}

\author{
\alignauthor Dan Olteanu\\[.5em]
       \affaddr{Department of Informatics, University of Zurich\\
       \email{olteanu@ifi.uzh.ch}\\[.5em]
       \url{https://fdbresearch.github.io}\hspace*{1em} \url{https://www.relational.ai}\\[.5em]
       \url{https://www.ifi.uzh.ch/en/dast.html}}
}

\date{}

\maketitle

\begin{abstract}
This paper overviews an approach that addresses machine learning over relational data as a database problem. This is justified by two observations. First, the input to the learning task is commonly the result of a feature extraction query over the relational data. Second, the learning task requires the computation of group-by aggregates. 

This approach has been already investigated for a number of supervised and unsupervised learning tasks, including: ridge linear regression, factorisation machines, support vector machines, decision trees, principal component analysis, and k-means; and also for linear algebra over data matrices.

The main message of this work is that the runtime performance of machine learning can be dramatically boosted by a toolbox of techniques that exploit the knowledge of the underlying data. This includes theoretical development on the algebraic, combinatorial, and statistical structure of relational data processing and systems development on code specialisation, low-level computation sharing, and parallelisation. These techniques aim at lowering both the complexity and the constant factors of the learning time.

\vspace*{1em}

This work is the outcome of extensive collaboration of the author with colleagues from RelationalAI, in particular Mahmoud Abo Khamis, Molham Aref, Hung Ngo, and XuanLong Ngu\-yen, and from the FDB research project, in particular Ahmet Kara, Milos Nikolic, Maximilian Schleich, Amir Shaikhha, Jakub Z{\'{a}}vodn{\'{y}}, and Haozhe Zhang. The author would also like to thank the members of the FDB project for the figures and examples used in this paper. 

The author is grateful for support from industry: Amazon Web Services, Google, Infor, LogicBlox, Microsoft Azure, RelationalAI; and from the funding agencies EPSRC and ERC. This project has received funding from the European Union's Horizon 2020 research and innovation programme under grant agreement No 682588.

This paper accompanies the keynote given at VLDB 2020 on September 1, 2020.

\end{abstract}

\section{Why Bother?}
\label{sec:introduction}

\begin{figure}
\centering
	  \includegraphics[scale=0.15]{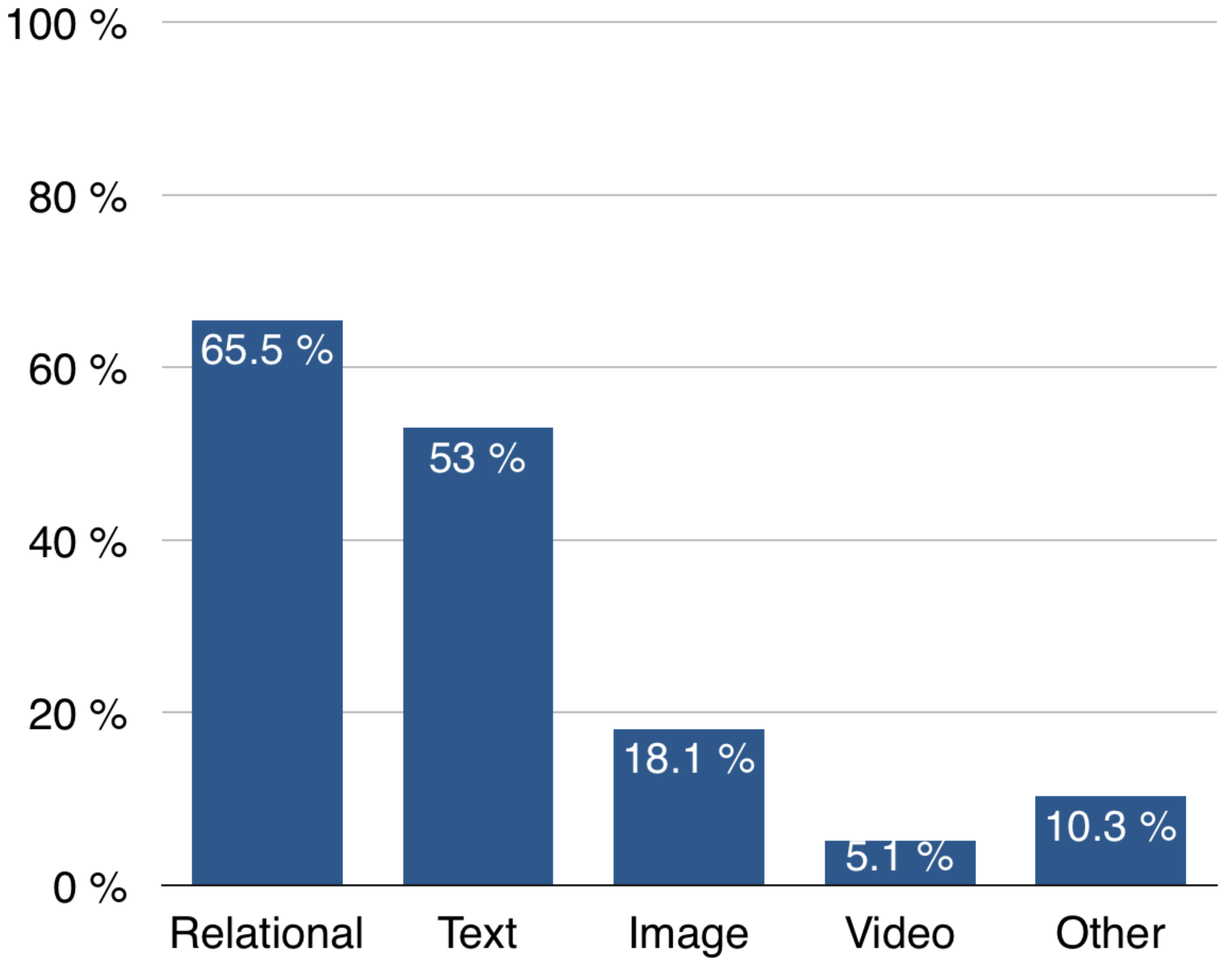}
  \hspace*{1em}
  \includegraphics[scale=0.15]{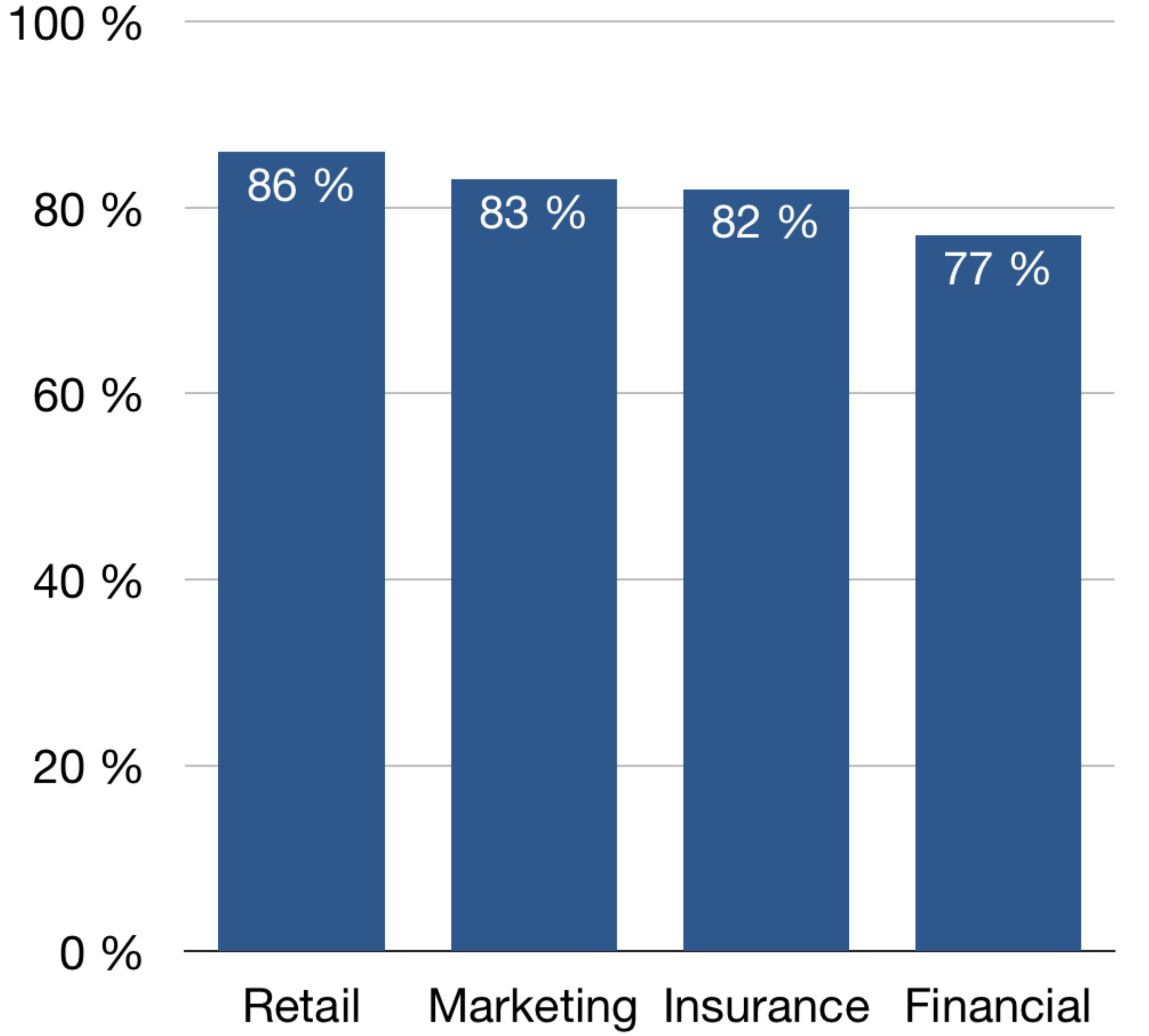}
	\label{fig:kaggle}
  \caption{Kaggle survey of 16,000 practitioners on the state of data science and machine learning: Use of relational data: overall (left), by industry (right).}
\end{figure}

As we witness the data science revolution, each research community legitimately reflects on its relevance and place in this new landscape. The database research community has at least three reasons to feel empowered by this revolution. This has to do with the pervasiveness of relational data in data science, the wide\-spread need for efficient data processing, and the new processing challenges posed by data science workloads beyond the classical database workloads. The first two aforementioned reasons are widely acknowledged as core to the community's {\em raison d'$\hat{e}$tre}. The third reason explains the longevity of relational database management systems success: Whenever a new promising data-centric technology surfaces, research is under way to show that it can be captured naturally by variations or extensions of the existing relational techniques. Prime examples are the management of XML data~\cite{XPath:SIGMOD:2002}, probabilistic data~\cite{PDB:Book:2011}, graph data~\cite{LB:SIGMOD:2015}, code~\cite{Semmle:ECOOP:2016}, and ontologies~\cite{Ontology:IJCAI:2018}, as well as prior work on migrating data mining inside the database~\cite{Chaudhuri:DMDB:1998}.

\begin{quote}\em
	Like the Star Trek's Borg Collective co-opting technology and knowledge of alien species, the Relational Data Borg a.k.a.\@ RDBMS assimilates ideas and applications from connex fields to adapt to new requirements and become ever more powerful and versatile. Unlike the former, the latter moves fast, has great skin complexion, and is reasonably happy. Resistance {\em is} futile in either case.
\end{quote}

\begin{figure*}
	\centering
	 \begin{tikzpicture}[scale=1.3]
   \node at (0,0) {{ }};

   \begin{scope}[shift={(0,-1.5)},scale=.8,every node/.style={transform shape, scale=.9}]


       \node at (-0.2,0.5) (inv) {};
       \maketableo{1}{1.5}{4}{10}{black}{-0.2}{0.5}{dgreen};
       \node at (-0.2,1.4) {\textbf{Inventory}};

       \node at (-1.4,0) (store) {};
       \maketableo{0.5}{0.6}{4}{5}{black}{-1.4}{0}{dgreen};
       \node  at (-1.4,0.45) {\textbf{Stores}};
       \draw ($(inv)+(-0.5,-0.1)$) -- ++(-0.1,0) |- ($(store)+(0.25,0)$);

       \node at (0.4,-0.9) (item) {};
       \maketableo{0.5}{0.6}{4}{5}{black}{0.4}{-0.9}{dgreen};
       \node at (0.4,-0.45) {\textbf{Items}};
       \draw ($(inv)+(0.1,-0.75)$) |- ($(item)+(-0.25,0)$);


       \node at (1.2,1) (weather) {};
       \maketableo{0.5}{0.6}{4}{5}{black}{1.2}{0.25}{dgreen}
       \node at (1.2,0.75) {\textbf{Weather}};
       \draw ($(inv)+(0.5,-0.1)$) |- ($(weather)+(-0.25,-0.6)$);


       \node at (-0.7,-0.8) (demo) {};
       \maketableo{0.5}{0.6}{4}{5}{black}{-0.7}{-0.8}{dgreen};
       \node[scale=0.9]  at (-0.7,-1.3) {\textbf{Demographics}};
       \draw ($(store)+(0,-0.3)$)  |- ($(demo)+(-0.25,0)$);
	
     \draw[thick,->,double] (2.5,1) -- (6.5,1) node[midway,scale=1,below, align=center]
     {Inventory $\Join$ Stores $\Join$ Items \\
       $\Join$ Weather $\Join$  Demographics };

     \draw[thick, ->, double, dred] (2.5,1) -- (6.5,1) node[midway,scale=1,above]
     {Feature Extraction Query};

    \end{scope}

   \begin{scope}[shift={(7,-1.5)}, scale=.8, every node/.style={transform shape}]
       \node[align=center] at (-0.1,1.8) {10,000s of Features};
       \maketableo{2.6}{3.0}{10}{60}{dred}{0}{0.1}{dred};
   \end{scope}


     \node[data] at (0,-3) (data) {Relational Data};
     \node[data] at (7,-3) (td) {Training Dataset};
     \node[data,scale=1] at ($(7,-4)$) (ML) {ML Tool};

    \draw[path, dred] ($(ML)+(0,.75)$) -- (ML);
    \draw[path, dred] (ML) -- ($(ML)+(0,-.75)$);

      \begin{scope}[shift={($(7,-6)$)}, scale=1.2]

        \draw [thick, <->] (0,1) -- (0,0) -- (1,0);
        \draw [oxfordblue,fill=oxfordblue] (.2,.2) circle [radius=0.025];
        \draw [oxfordblue,fill=oxfordblue] (.3,.2) circle [radius=0.025];
        \draw [oxfordblue,fill=oxfordblue] (.2,.3) circle [radius=0.025];
        \draw [oxfordblue,fill=oxfordblue] (.3,.4) circle [radius=0.025];
        \draw [oxfordblue,fill=oxfordblue] (.6,.5) circle [radius=0.025];
        \draw [oxfordblue,fill=oxfordblue] (.7,.5) circle [radius=0.025];
        \draw [oxfordblue,fill=oxfordblue] (.4,.6) circle [radius=0.025];

        \draw [dred,thick] (0.1,0.05) -- (0.8,0.8);
      \end{scope}
      
     \node[data,scale=1, fill=white] at ($(7.6,-5.9)$) (model) {Model};
         
      \begin{scope}[shift={($(-1.5,-6)$)}]
        \foreach \i in {13,...,1}
        \maketableo{0.25}{0.3}{2}{3}{black}{\i/10}{\i/10}{oxfordblue};
        \foreach \i in {13,...,1}
        \maketableo{0.25}{0.3}{2}{3}{black}{\i/10+.6}{\i/10}{oxfordblue};      
        \foreach \i in {13,...,1}
        \maketableo{0.25}{0.3}{2}{3}{black}{\i/10+1.2}{\i/10}{oxfordblue};      
        \foreach \i in {13,...,1}
        \maketableo{0.25}{0.3}{2}{3}{black}{\i/10+1.8}{\i/10}{oxfordblue};      
        \foreach \i in {13,...,1}
        \maketableo{0.25}{0.3}{2}{3}{black}{\i/10+2.4}{\i/10}{oxfordblue};      
      \end{scope}

    \node[data, oxfordblue, fill=white, align=center]  at (0.2,-5.75) (batch) {Batch of \\Aggregate Queries};

    \draw[path, thick, oxfordblue] (data.south) -- ++(0,-.9);
    
      \node[data, oxfordblue, scale=1] at($(batch)+(4.2,0)$) (GD)
      {Optimisation\\\vspace{.1cm}};
      \begin{scope}[shift={($(GD)+(0,-.2)$)},scale=.2,every node/.style={transform shape}]
        \draw[oxfordblue,->,thick] (-1,.1) arc (170:10:1);
        \draw[oxfordblue,->,thick] (1,-.1) arc (-10:-170:1);
      \end{scope}
      \draw[oxfordblue, thick, path, <->] ($(batch.east)+(0.4,0)$) -- ++(1.5,0);

    \draw[oxfordblue, path, thick] (GD.east) -- ++(1.4,0);

    \node[oxfordblue] at ($(data)+(1.5,-0.5)$)
    {Feature Extraction Query};
    \node[oxfordblue] at ($(data)+(1.5,-0.8)$) 
    {+ Feature Aggregates };

 \end{tikzpicture}

\caption{Learning over relational data. In {\color{dred}structure-agnostic learning} (the top flow in red), a feature extraction query constructs the data matrix on which the model is trained using a  machine learning library. In {\color{oxfordblue}structure-aware learning} (the bottom flow in blue), sufficient statistics is computed over the input database using a batch of aggregate queries and then optimisation is performed on this statistics to obtain the model.}
\label{fig:state-of-affairs}
\end{figure*}
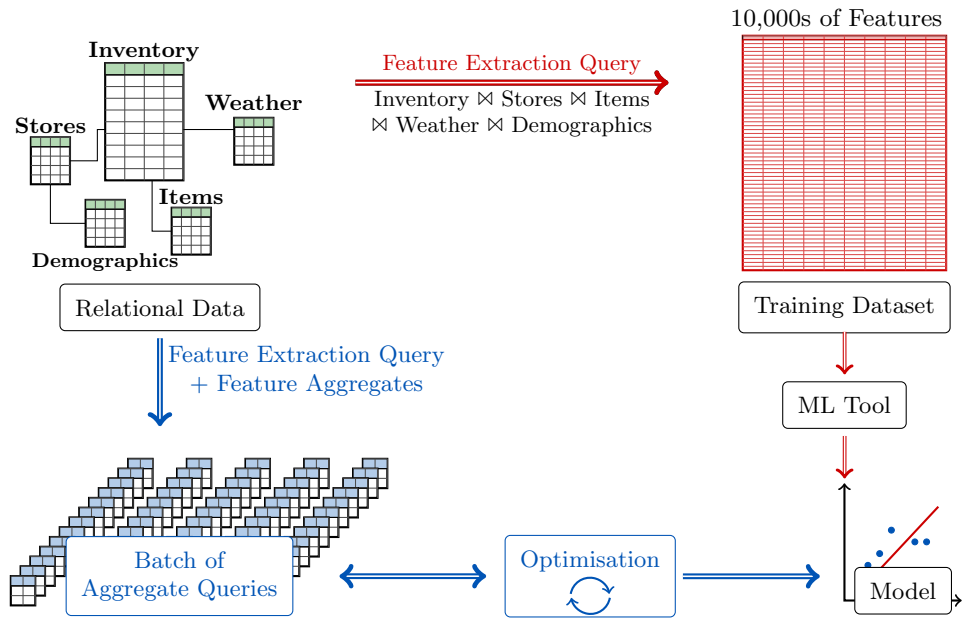

This paper overviews work by the author on learning over relational data. Along with a growing number of recent contributions, e.g., a database theory manifest~\cite{DBTHEORY:DAGSTUHL:17-small} and SIGMOD tutorials overviewing work at the interface of machine learning and databases~\cite{Polyzotis:SIGMOD:Tutorial:17,Kumar:SIGMOD:Tutorial:17}, this effort is another episode of the relentless learning process for the Relational Data Borg.

\subsection{Relational data is ubiquitous}

\begin{quote}\em According to a recent Kaggle survey~\cite{kaggle-survey}, most data scientists use relational data  (Figure~\ref{fig:kaggle}).	
\end{quote}

This is a testament to the usability of the relational data model. Many human hours are invested in building relational databases that are curated and enriched with knowledge of the underlying domains and integrity constraints.

Figure~\ref{fig:state-of-affairs} depicts a dataset used for data science in the retailer domain. Features are gathered from a number of relations detailing information on items in stores, store information, demographics for areas around the stores, inventory units for items in stores on particular dates, and weather.
The data matrix, which is the input to the learning task, is the result of a {\em feature extraction query} that joins these relations on keys for dates, locations, zipcode, and items. The query may construct further features using aggregates, e.g., running aggregates over days, weeks, months; min, max, average, median aggregates, or aggregates over many-to-many relationships and categorical attributes with high cardinality (e.g., identifiers)~\cite{Perlich:ML:2006}. A typical model of interest would predict sales or inventory units for next month.

The author's observation based on interactions with data scientists at LogicBlox and RelationalAI is that similar data\-sets of up to a few hundreds of millions of rows across a dozen relations are  common in data science projects.

\nop{Categorical variables such as day-of-year or item identifier are typically one-hot encoded, leading to wide feature vectors, which are the rows in the matrix.} 

\subsection{Ignorance is not bliss}
\label{sec:stateofaffairs}

\begin{quote}\em 
	The mainstream data science solutions ignore the structure of the underlying relational data at the expense of runtime performance.
\end{quote}

\begin{figure*}
\begin{tabular}{lrrc}\toprule
    {\bf Relation} & {\bf Cardinality/Attrs} & {\bf CSV Size} \\\midrule
    Inventory      & 84,055,817 / \hspace*{.5em}4                & 2 GB                  \\
    Items          & 5,618 / \hspace*{.5em}5                & 129 KB                \\
    Stores         & 1,317  / 15               & 139 KB                \\
    Demogr.        & 1,302  / 16               & 161 KB                \\
    Weather        & 1,159,457  / \hspace*{.5em}8                & 33 MB                 \\\midrule
    Join           & 84,055,817 / 44               & 23GB                  \\\bottomrule
  \end{tabular}\hspace*{1em}%
\begin{tabular}{lrrrr}\toprule
 & \multicolumn{2}{c}{{\color{dred}\bf PostgreSQL+TensorFlow}} & \multicolumn{2}{c}{{\color{oxfordblue}\bf LMFAO}~\cite{lmfao}} \\
           & {\bf Time}  & {\bf CSV Size} & {\bf Time}       & {\bf CSV Size} \\\midrule
Database   & --          & 2.1 GB     & --       & 2.1 GB    \\
Join       & 152.06 secs    & {\color{dred}23 GB}      & --         & -- \\
Export     & 351.76 secs    & {\color{dred}23 GB}      & --         & -- \\
Shuffling  & 5,488.73 secs  & {\color{dred}23 GB}      & --         & -- \\ 
Query batch & --          & --         & 6.08 secs    & {\color{oxfordblue}37 KB} \\
Grad Descent         & 7,249.58 secs& --         & 0.05 secs   & -- \\\midrule
Total time      & {\color{dred}13,242.13 secs} &          & {\color{oxfordblue}6.13 secs}  & \\\bottomrule
\end{tabular}
	\caption{Left: characteristics of our retailer dataset. Right: Runtime performance experiment (Intel i7-4770, 3.4GHz, 32GB, 8 cores). The {\color{oxfordblue}structure-aware (LMFAO)} approach is 2,160x faster than the {\color{dred}structure-agnostic (PostgreSQL+TensorFlow)} approach, while producing a more accurate solution (RMSE on 2\% test data). TensorFlow runs one epoch with 100K tuple batch.
		Both approaches train a linear regression model to predict the inventory given all the other features from the data matrix defined by the join of the relations.	}
	\label{fig:comparison}
\end{figure*}

The structure of relational data is rich. A large body of work in database research is on discovering, enforcing, and exploiting structure such as dependencies and integrity constraints. It is therefore rather perplexing that the current state of affairs in learning over relational data is to ignore its structure! The flow of such {\color{dred}structure-agnostic} mainstream solutions is depicted in Figure~\ref{fig:state-of-affairs}: They first construct the data matrix using a feature extraction query expressed in SQL for a database system, e.g., PostgreSQL or SparkSQL~\cite{Spark:NSDI:2012}, or in Python Pandas~\cite{pandas} for jupyter notebooks. The data matrix is then passed on to a machine learning library, e.g., scikit-learn~\cite{scikit2011-small}, R~\cite{R-project}, TensorFlow~\cite{tensorflow-small}, or MLlib~\cite{MLlib:JMLR:2016-small}, which then learns the model. 

The structure-agnostic solutions that put together black-box specialised systems for data processing and machine learning may work for virtually any dataset and model. The uncomfortable drawback of this marriage of convenience is that the two systems were not originally designed to work together. They may therefore suffer from some of the following shortcomings: (1) the feature extraction query is fully materialised; (2) its result is exported from the query engine and imported into the statistical library; (3) the categorical variables are one-hot encoded; (4) the pipeline may suffer from impedance mismatch and high maintenance cost in the face of changes to the underlying dataset; (5) and the pipeline inherits the limitations of both systems.

In practice, all of these shortcomings may significantly hinder the runtime performance of the data science solution. 

\medskip

{\noindent\bf 1. Data matrix materialisation.}  The materialisation of the feature extraction query may take non-trivial time as it typically produces data matrices whose sizes are orders of magnitude larger than the input data. This departs from the classical database setting, where queries are over-constrained by highly selective joins and filter conditions. Here, the goal is to bring together the available data and use it to train accurate models. Figure~\ref{fig:comparison} shows that for our retailer dataset, the result of the key-fkey join is one order of magnitude larger than the input data\footnote{A notable exception is a factorised query engine~\cite{FDB:PVLDB:2012,BKOZ13,lmfao}. For our retailer dataset, the factorised/non-factorised join is 26x smaller/10x larger than the input data size.}.

\medskip

{\noindent\bf 2. Data move.} Moving the data matrix from the query engine to the learning library is without doubt a time-consu\-ming step. Furthermore, the data formats of the two systems may be different and require a conversion. In our experiment reported in Figure~\ref{fig:comparison}, moving the data matrix takes more than twice the time to compute it.

\medskip

{\noindent\bf 3. One-hot encoding.} The data is standardised and one-hot encoded before the learning step. Done na\"ively, the one-hot encoding may increase the size of the data matrix, turning it from lean (tall-and-thin) into chubby and thereby blurring the typical database distinction between the large number of rows and the small number of columns.

\medskip

{\noindent\bf 4. Maintenance cost.} The structure-agnostic data science pipeline is not designed to cope effectively with on-the-fly changes to the underlying data. Data updates in the form of tuple insertions and deletions, as well as schema updates in the form of addition and removal of relations, columns, derived aggregate features, would require recomputation of the data matrix as well as the model learned over it.

\medskip

{\noindent\bf 5. Inherited limitations.} The two systems have been designed for different use cases that are not necessarily aligned. For instance, the maximum data frame size in R and the maximum number of columns in PostgreSQL are much less than typical database sizes and respectively number of model features. The mainstream solutions thus inherit the limitations of both constituent systems.

All aforementioned shortcomings pose significant runtime performance challenges. Whereas query processing engines are designed to cope with large datasets, this is not the case for the bulk of popular learning libraries. By having the learning libraries work on orders of magnitude larger data than the query engine, the mainstream solutions exacerbate the runtime performance challenge. Even for medium-sized input datasets such as our retailer dataset, working solutions are forced to employ distributed platforms to train models. This accidental complexity comes with a high, rather unnecessary compute cost, as observed repeatedly in recent studies on the cost of scalability~\cite{ScalabilityCost:HOTOS:2015}, and also non-trivial energy cost~\cite{ComputingImpact:CACM:2019}. Distributed platforms should be reserved for truly very large problem instances to justify the cost.

\subsection{Structure awareness to the rescue!}

\begin{quote}
 {\em Learning over the data matrix defined by a feature extraction query may take much less time than materialising the data matrix and is feasible on a commodity machine.}
 \end{quote}

A natural question is whether the shortcomings highligh\-ted in Section~\ref{sec:stateofaffairs} can be avoided. There is increasing effort to address these shortcomings by migrating from the mainstream solution to an ever tighter integration of the querying and the learning tasks into a single execution plan that avoids moving the data matrix. Opening up the learning task and expressing its data-intensive computation steps as a {\em batch of aggregate queries} is another source of radical performance improvements. Computing the batch of queries over the feature extraction queries is now a purely database workload. We can exploit the join structure in the feature extraction query to optimise this database workload. In fact, an entire toolbox of database theory and systems techniques is at our disposal to tackle the performance challenges of this workload, as discussed in the next sections. This forms the basis of the {\color{oxfordblue}structure-aware learning} paradigm, as opposed to the {\color{clr_outofDB}structure-agnostic} paradigm discussed before. 

Figure~\ref{fig:state-of-affairs} depicts the flow of a structure-aware learning solution: By analysing the learning task, i.e., the features of the model and the objective function used for learning the model, a batch of aggregates is synthesised manually~\cite{SOC:SIGMOD:16} or automatically~\cite{IFAQ:CGO:2020}. This batch is then composed with the feature extraction query, optimised, and evaluated. Its result is typically much smaller than that of the feature extraction query and its evaluation may be much faster than that of the feature extraction query alone! This result is sufficient to compute the model parameters using an optimisation algorithm tailored at the desired model such as gradient descent or CART. Section~\ref{sec:aggregates} exemplifies aggregates for several models and objective functions.

\begin{figure*}
	\centering
	\hspace*{-1em}\includegraphics[scale=.325]{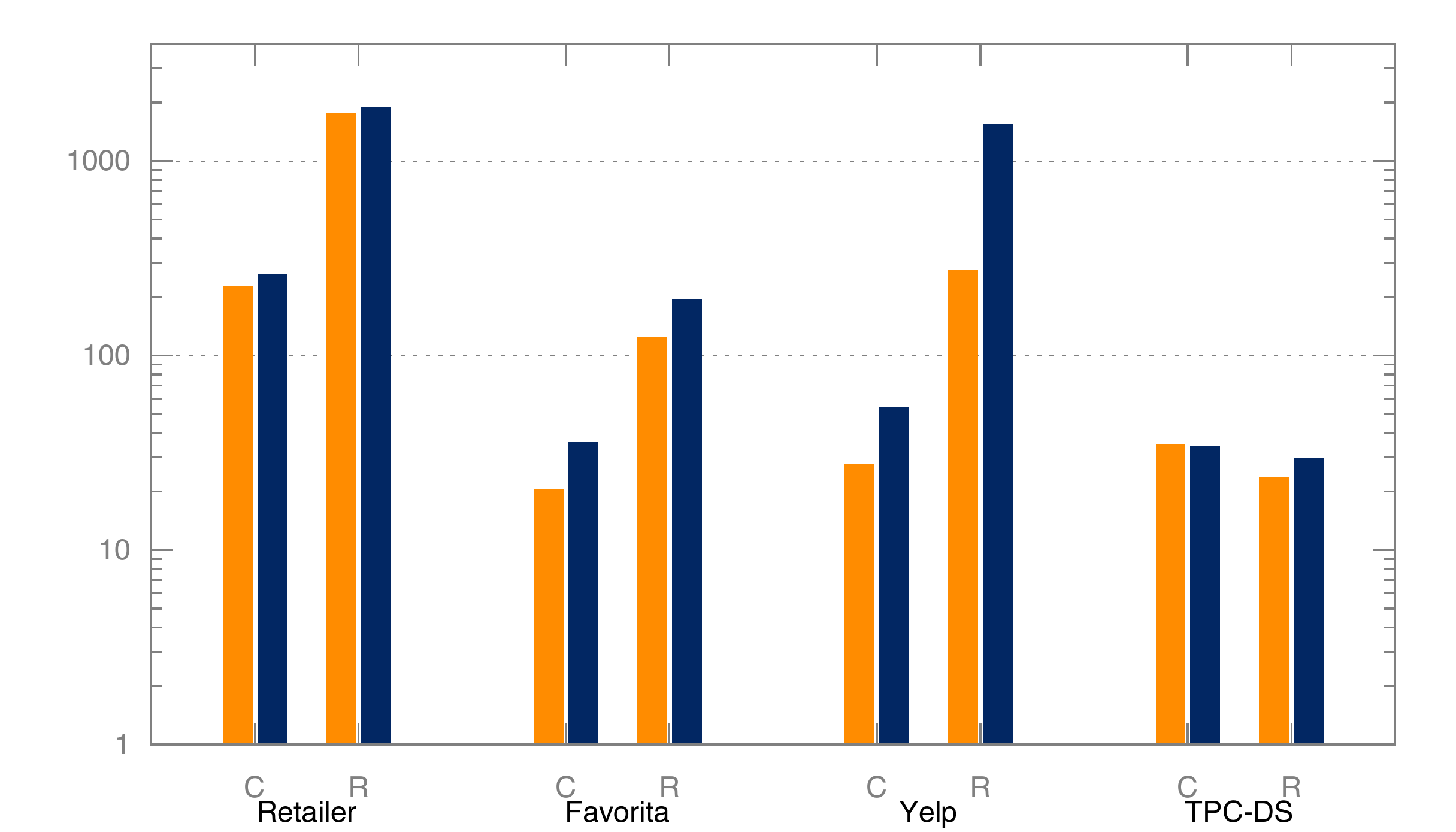}
    \includegraphics[scale=.55]{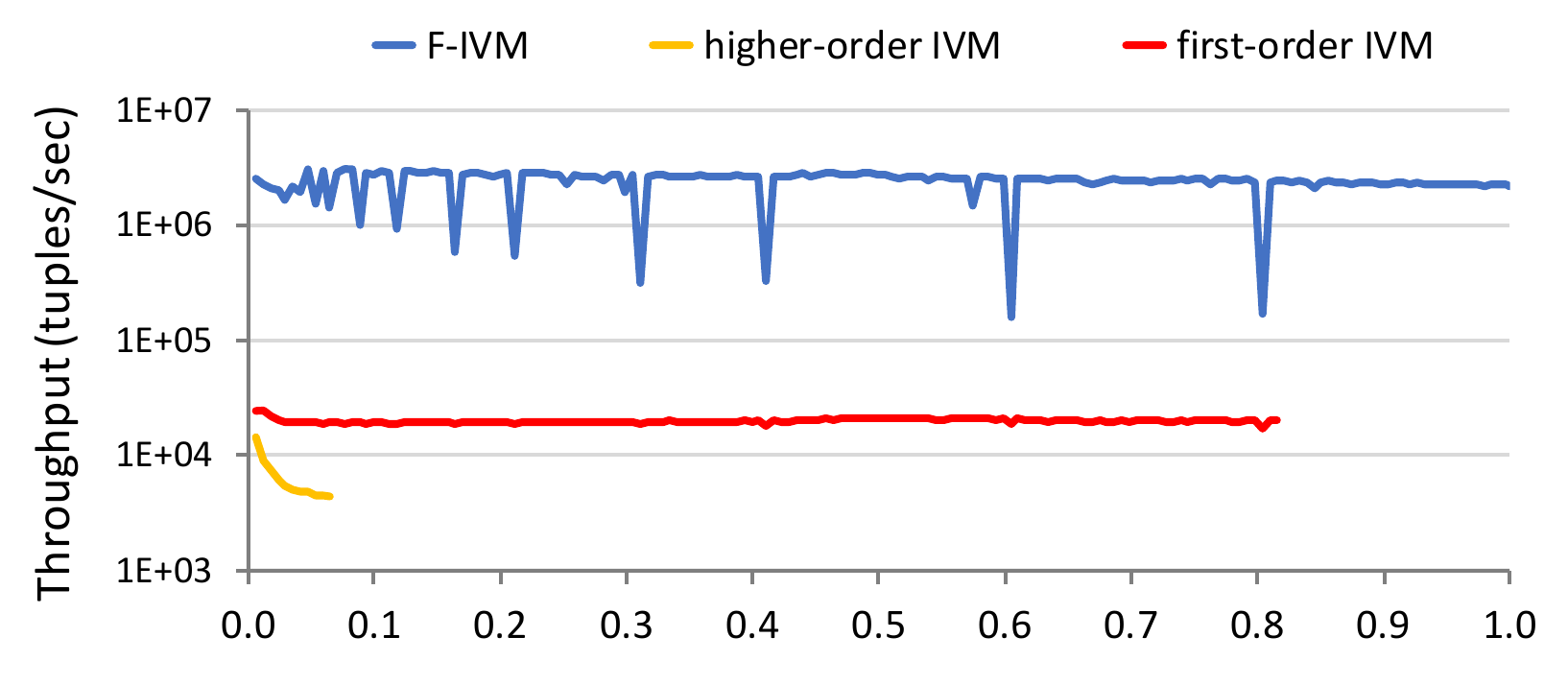}
    \caption{Left: Speedup of LMFAO over existing DBMSs (commercial {\color{darkorange}\bf DBX} and {\color{dblue}\bf MonetDB}). The task is to compute query batches C (covariance matrix) and R (regression tree node) over four datasets (AWS d2.xlarge, 4 vCPUs, 32GB). Right: Throughput of F-IVM over first-order IVM (delta processing) and higher-order IVM (delta processing with intermediate views). The task is to maintain the covariance matrix in the retailer dataset (Azure DS14, Intel Xeon, 2.40 GHz, 112GB, 1 thread, one hour timeout).}
    \label{fig:existing-technology}
\end{figure*}

Our structure-aware learning system LMFAO (Layered Multiple Functional Aggregates Optimisation)~\cite{lmfao} avoids the materialisation of the join and computes the batch of aggregates directly over the input database. As shown in Figure~\ref{fig:comparison}, LMFAO only takes 6 seconds to compute the batch. It then takes 50 milliseconds to compute the model parameters of a ridge linear regression model using gradient descent, where the gradient vector is built up using the computed aggregates and the current parameters. 

In contrast, the structure-agnostic solution takes significantly more time. It needs 152 seconds only to compute the join using PostgreSQL. While commercial database systems may be faster than PostgreSQL, they would still need non-trivial time to create the 23GB join result, whereas the sufficient statistics computed by LMFAO is only 37KB. The structure-agnostic solution exports the data matrix from PostgreSQL, imports it into TensorFlow, and shuffles it. Finally, the model is learned using a variant of stochastic gradient descent in one epoch (i.e., one pass over the data matrix). Both times to shuffle and learn using TensorFlow dwarf the join time by a factor of 50x. Overall, the structure-aware solution is 2,160x faster than the structure-agnostic solution in our experiment. Both systems were given the same set of input features and the resulting model was validated against a training dataset. LMFAO computed a model very close to that given by the closed-form ordinary least squares solution, whereas TensorFlow returned a slightly less accurate model trained in one epoch (pass) over the data matrix. 

This lack of efficiency can be observed beyond the choice of model and dataset in our experiment. Prior work~\cite{SOC:SIGMOD:16,ANNOS:DEEM:18,lmfao} reports on similar stories with a range of public datasets (Favorita, TPC-DS, Yelp), models (decision trees, factorisation machines, k-means), and learning libraries (R~\cite{R-project}, scikit-learn~\cite{scikit2011-small}, Python StatsModels~\cite{P-StatsModels}, mlpack~\cite{mlpack2018-small}, and XGBoost~\cite{xgboost}). They often fail to process the data matrix on commodity machines due to out-of-memory errors. 

Over the past years, we have investigated the gap between the two types of solutions from different angles and for different models. We have built a number of prototypes for learning a variety of models over relational data: F~\cite{SOC:SIGMOD:16,OS:PVLDB:2016}, F-IVM\footnote{\url{https://github.com/fdbresearch/FIVM}}~\cite{Nikolic:FIVM:2018,FIVM:SIGMOD:2020}, AC/DC~\cite{ANNOS:DEEM:18}, LMFAO~\cite{lmfao,LMFAO:PVLDB:2020}, IFAQ~\cite{IFAQ:CGO:2020}, and Rk-means~\cite{Rkmeans:AISTATS:2020}. We have also investigated theoretical aspects including: the use of sparse tensors and functional dependencies to improve the time complexity of learning polynomial regression models, factorisation machines, and PCA over relational data~\cite{ANNOS:PODS:2018,ANNOS:TODS:2020}; algorithms and complexity analysis for queries with group-by aggregates and joins with equality and additive inequality conditions, and their application to k-means clustering and learning linear support vector machines and models with non-polynomial loss~\cite{faqai}; and the incremental maintenance of linear regression models~\cite{Nikolic:FIVM:2018} and of relational queries~\cite{IVMe:ICDT:2019,IVMe:PODS:2020} under data updates. Sections~\ref{sec:asymptotics}, \ref{sec:constants}, and \ref{sec:highlights} highlight some of this development.

Our effort is not alone. There is an increasingly larger body of work at the interface of databases and machine learning towards addressing the shortcomings listed in Section~\ref{sec:stateofaffairs}. For instance, increasingly more solutions, starting with MADlib~\cite{MADlib:2012-small} and Bismarck~\cite{Kumar:InDBMS:2012}, integrate the two systems into one system to avoid Shortcomings (2), (4), and (5). This mirrors two-decades old efforts on integrating data mining and databases~\cite{Chaudhuri:DMDB:1998}. Shortcoming (3) is commonly addressed by compressed feature vectors that avoid explicit representation of zero values as done by, e.g., TensorFlow~\cite{tensorflow-small} and libFM~\cite{libfm}. Our previous work~\cite{ANNOS:PODS:2018,ANNOS:TODS:2020} proposed a sparse tensor representation for one-hot encoded features that can be expressed using group-by queries; this is used by LMFAO~\cite{lmfao}. A few solutions aim at addressing Shortcoming (1) by pushing the learning task past the joins. Whereas most solutions  only handle restricted joins, e.g., star key-fkey joins~\cite{KuNaPa15}, others allow for arbitrary join schemas~\cite{SOC:SIGMOD:16,lmfao}. A brief and incomplete overview of such approaches is given in a recent tutorial~\cite{SOANN:SUM:2019}. A promising area of research is understanding the connection between learning and query languages, e.g., the relative expressiveness of graph neural networks and fragments of first-order logic and SQL~\cite{GNNQL:SIGREC:2020}.

\subsection{Wait! This problem was already solved}

\begin{quote}
	{\em Death dealers, who categorise research problems as either "done/dead" or "not done", might say that the problem of learning over relational data is done. Yet existing database technology is not ready for workloads specific to training models.}
\end{quote}

Given that the structure-aware solution essentially relies on casting data-dependent computation as queries, why not use existing database systems to compute them? 

Figure~\ref{fig:existing-technology} depicts two of our experimental findings. 

The first experiment considers the computation of the aggregates that define the covariance matrix used to learn a linear regression model and the costs used by CART to decide on a node in a regression tree~\cite{lmfao}. The number of distinct aggregates varies from hundred to thousands for the four considered datasets, yet they are very similar and amenable to significant shared computation. The speedup of our prototype LMFAO over two competitive data\-base systems is on par with the number of aggregates, suggesting that these systems might only provide limited support for sharing computation across the aggregates in a batch. These systems performed worse on SQL encodings of LMFAO's shared execution plans, which compute many intermediate aggregates. This hints at LMFAO’s distinct design that departs from mainstream query processing.

The second experiment considers the maintenance of the covariance matrix under tuple insertions into an initially empty retailer dataset~\cite{Nikolic:FIVM:2018}. The throughput of our system F-IVM (Factorised Incremental View Maintenance) is orders of magnitude higher than of first-order (classical delta processing) and higher-order (delta processing with intermediate views) incremental view maintenance approaches. The three approaches are implemented on top of DBToaster's backend~\cite{Koch:DBToaster:2014} and take advantage of its intermediate representation language and code compilation. The main reason for the performance difference lies again in sharing the maintenance across the aggregates within a batch.

These experiments suggest that the task of learning over relational data comes with new database workloads that require novel optimisation and evaluation techniques. LMFAO and F-IVM are just two points of reference in a large design space for systems to cope with the new workloads.

\subsection{Wait, wait! This effort is irrelevant}

\begin{quote}
	{\em This effort misses the point: Accuracy is the holy grail of machine learning, not runtime efficiency. Yet faster training can mean better accuracy.}
\end{quote}

The term ``performance" is mostly used to mean accuracy in the machine learning literature and runtime efficiency in the database community. How can a database mindset ever make a difference to the machine learning battle field? It is obviously pointless to quickly train models with poor accuracy and therefore not useful. While this is a valid point, being able to train much faster can in fact help improve the accuracy in several scenarios.

 First, it raises the possibility to train several models within the same time that it takes a slower system to train one of these models. For instance, in our experiment in Figure~\ref{fig:comparison}, once we computed the covariance matrix over the given features, we can train a new model over a subset of these features in 50 milliseconds. In contrast, TensorFlow would use separate scans of the data matrix to train each model, with each scan taking over 7,000 seconds. This would effectively allow the faster approach to perform model selection by looking over a large set of possible models defined by subsets of the set of all given features within a fraction of the time it takes TensorFlow to complete one model. 
 
Second, it allows to train more complex and expensive models in a reasonable time.

Third, it opens the door to keeping models fresh so that they accurately reflect the new evidence: Instead of updating the models say every day solely due to the abysmal runtime performance of the learning library, one can update it every hour or even every minute. In our streaming experiment in Figure~\ref{fig:existing-technology} (right), F-IVM can maintain the covariance matrix with a throughput of over one million tuples per second. Refreshing the linear regression model after each update (we considered a bulk of 1000 inserts at a time) would take  tens of milliseconds only; it takes less than the 50 milliseconds for computing the parameters from scratch given the covariance matrix, since we resume the convergence procedure of the gradient descent already with parameter values that are close to the final ones.

Besides the runtime aspect, the structure of the relational data and the human-added knowledge in the form of constraints and ontologies may help synthesise more accurate models. This is a relevant subject that is out of the scope of this paper and deserves proper treatment on its own.

\nop{

Key Argument 1: speeding up ML
\begin{itemize}
		\item state of the art: holistic combinations of full-fledged DBMS + full-fledged ML library.
		\item no-brainer 1: reducing data movement. MADlib etc.
		\item harder but promising approach that we advocate for in this manifesto: tight integration of ML and DB -- a slow burner, requires non-trivial work, not ready-made meal, but intellectually rewarding and practically relevant
		\item we are good at improving runtime performance. ML folks are however interested mostly in accuracy (the term "performance" is used for accuracy in ML community and runtime in DB community). Where can we contribute? improving runtime performance can have several impacts: enumerate them here
		\item Example: we can be non-trivially faster than staet of the art solutions, show the case of Retailer.
		\item Also, expert knowledge in DB modeling can boost accuracy of models. LB did that. any citation? this should be pursued more.
\end{itemize}

Key Argument 2: New DB setting to explore
\begin{itemize}
	\item we have beaten the path of query processing for so long. anything new? the doomsayers in the community say: all is done/solved. but nothing is really truly done in research. new setting: queries not selective, optimization not as much about join ordering due to selectivities of predicates, but to the large amount of group-by aggregates to be computed over the same join. existing DB technology does not seem

	\item Example 2: How to DBMSs and DSMSs fare for such workloads?
\end{itemize}
}

\begin{figure}
\begin{center}
\begin{small}
\begin{tabular}{lrrrr}\toprule
  & {\bf Retailer}
  & {\bf Favorita}
  & {\bf Yelp}  
  & {\bf TPC-DS} \\\midrule
  Covar. matrix
  & 937
  & 157
  & 730
  & 3,299 \\
  Decision node
  & 3,150
  & 273 
  & 1,392
  & 4,299 \\
  Mutual inf.
  & 56 
  & 106 
  & 172 
  & 254  \\
  $k$-means
  & 44
  & 19
  & 38 
  & 92\\\bottomrule
\end{tabular}
\end{small}
\end{center}
\caption{Number of aggregates for various datasets and workloads (covariance matrix for linear regression; decision tree node; mutual information for model selection and Chow-Liu trees; $k$-means).}
\label{fig:aggregates}
\end{figure}

\section{From Learning to Aggregates}
\label{sec:aggregates}

\begin{quote}
	{\em Aggregation is the aspirin to all problems.}
\end{quote}

As argued in Section~\ref{sec:introduction}, turning the machine learning problem into a database problem  follows a databa\-se community's recipe for success. The learning task relies on various forms of aggregation over the data matrix that are readily supported by database query languages. The types of aggregates depend on the model and objective function. In this section, we give such aggregates for several models; it is beyond the scope of this overview paper to show how to derive them. For an in-depth treatment, please refer to our work~\cite{ANNOS:PODS:2018,lmfao,ANNOS:TODS:2020,faqai}. Figure~\ref{fig:aggregates} shows that their number is much more than in a typical database query. The efficient computation of batches of such aggregates requires new techniques that exploit their structure and similarities. 

 We next consider that the data matrix $D$, which is defined by a feature extraction query, consists of tuples $(\mv x, y)$ of a feature vector $\mv x$ and a response $y$.

\subsection{Least-squares loss function}
\label{sec:least-squares-loss}

\begin{quote}
	{\em The gradient vector of the least-squares loss function is built up using sum-product aggregates over the model features and parameters.}
\end{quote}

For linear regression models trained by minimising the least-squares loss function using gradient descent optimisation, the data-dependent aggregates that come up in the gradient vector have one of the following forms:
\begin{align*}
	&\texttt{SUM}(X_i \texttt{*} X_j) & &\\
	&\texttt{SUM}(X_i) &\texttt{ GROUP BY } &X_j\\
	&\texttt{SUM}(1)   &\texttt{ GROUP BY } &X_i, X_j		
\end{align*}
where $X_i$ and $X_j$ are database attributes whose values make up the features $x_i$ and respectively $x_j$. In case both $x_i$ and $x_j$ are continuous, we sum their products over all tuples in the data matrix $D$. In case $x_i$ is continuous and $x_j$ is categorical, we report the sum of $x_i$ for each category of $x_j$. Finally, in case both $x_i$ and $x_j$ are categorical, we count the number of occurrences for each pair of their categories that appears in $D$. Each pair of features, or a pair of a feature and the response, defines such an aggregate; for $n$ features, we thus have $(n+1)^2$ such aggregates (although the aggregate $\texttt{SUM}(y^2)$ over the response $y$ is not needed in the gradient vector). We arrange these aggregates in an $(n+1)\times(n+1)$ symmetric matrix, where the $(i,j)$ entry is the aggregate for $X_i$ and $X_j$. If the features are standardised, this matrix is a compact encoding of the covariance matrix of the features.

Two remarks are in order. 

First, the data matrix $D$ need not be materialised, so the above aggregates are expressed directly over the feature extraction query $Q$:
\begin{align*}
	&\texttt{SELECT } X,  \texttt{agg FROM } Q \texttt{ GROUP BY } X;
\end{align*}
where $X$ represents the categorical feature(s) and \texttt{agg} is one of the sum aggregates discussed before. Parts of the aggregates can be pushed past the joins and shared across the aggregates. This reduces the sizes of the intermediate results. For the evaluation of such queries, we resort to factorised query processing~\cite{OlZa15,OS:SIGREC:2016} that combines aggregate pushdown with worst-case optimal join computation~\cite{LFTJ} to achieve a runtime asymptotically lower than classical query plans with binary joins~\cite{skew}. The feature extraction queries are typically acyclic, in which case our execution strategy takes time linear in the {\em input} data instead of output data.

Second, these aggregates represent a sparse tensor encoding of the interactions of any two categorical features: Instead of one-hot encoding them, we only represent the pairs of categories of the features that appear in the data matrix $D$~\cite{ANNOS:TODS:2020}. This is naturally expressed using the \texttt{group-by} clause.

Similar aggregates can be derived for polynomial regression models~\cite{OS:SIGREC:2016}, factorisation machines~\cite{ANNOS:PODS:2018}, sum-product networks~\cite{George:thesis:2020}, principal component analysis~\cite{ANNOS:PODS:2018}, quadratically regularised low-rank models~\cite{Gabriel:thesis:2019}, and QR and SVD decompositions~\cite{Bas:thesis:2018}.

\subsection{Cost functions for decision trees}

\begin{quote}
	{\em The computation of the cost functions for each attribute and condition at a decision tree node can be expressed by a sum-product aggregate with a filter condition.}
\end{quote}

The cost functions used by algorithms such as CART~\cite{cart84} for constructing regression trees rely on aggregates that compute the variance of the response conditioned on a filter:
\begin{align*}
\texttt{VARIANCE}(Y) \texttt{ WHERE } X_i \texttt{ op } c_j	
\end{align*}
For a continuous feature $x_i$, the filter asks whether its value is above a threshold $c_j$: $X_i \geq c_j$. For a categorical feature $x_i$, the filter asks whether its value is in a set of possible categories: $X_i \texttt{ in } (v_1,\ldots,v_k)$ or $X_i = v$. The thresholds and categories are decided in advance based on the distribution of values for $x_i$. The variance aggregate is expressed using the sum of squares, the square of sum, and the count.

For classification trees, the aggregates encode the entropy or the Gini index using group-by counts to compute value frequencies in the data matrix. 

The aggregates for decision trees are thus very similar to those in Section~\ref{sec:least-squares-loss}, with the addition of filters.

\medskip

\subsection{Non-polynomial loss functions}
\label{sec:nonpolynomial-loss}

\begin{quote}
	{\em The sub-gradients of non-polynomial loss functions require a new type of theta joins with additive inequality conditions. Such joins call for new algorithms beyond the classical ones.}
\end{quote}

A large class of models including support vector machines and robust regression are trained using sub-gradient descent. They use non-polynomial loss functions, such as (ordinal) hinge, scalene, and epsilon insensitive, that are defined by multiple cases conditioned on additive inequalities of the form $\sum_i x_i\cdot w_i > c$, where $w_i$ and $c$ are constants and $x_i$ are the features. Huber loss admits a gradient with additive inequalities. The (sub)gradients of these loss functions can be formulated using aggregates of the form~\cite{faqai}:
\begin{align*}
\texttt{SUM}(X) \texttt{ WHERE } X_1 * w_1 + \ldots + X_n * w_n	> c \texttt{ GROUP BY } Z
\end{align*}
where $X$ may be the value one, a feature, or the product of two attributes for continuous features, while $Z$ is none, one, or two attributes for categorical features.

This aggregate is structurally different from the previous ones as it involves an {\em additive inequality} join condition. This is a new type of theta join. Existing database systems evaluate it by iterating over the data matrix and checking the additive inequality condition for each tuple. As we show in recent work~\cite{faqai}, there are better algorithms that do not need to iterate over the entire data matrix and that may need polynomially less time to compute such aggregates with additive inequalities. This is a prime example of a new database workload motivated by a machine learning application.

Similar aggregates are derived for $k$-means clustering~\cite{faqai}.

\section{Lowering the Asymptotics}
\label{sec:asymptotics}

\begin{quote}
	{\em Over decades, the database community has developed a toolbox of techniques to understand and use the algebraic, combinatorial, statistical, and geometric structure of relational data to lower the computational complexity of database problems.} 
\end{quote}

In this section, we overview principles behind structure-aware learning that led to new algorithms for queries with joins, additive inequality conditions, and group-by aggregates, and the analysis of their computational complexity.

\subsection{Algebraic structure}

Relational data exhibits an algebraic structure: A relation is a sum-product expression, where the sum is the set union and the product is the Cartesian product. Relational algebra computation can be captured using (semi)rings\footnote{
A ring $(\RING, \RINGPLUS, \RINGPROD, \RINGZERO, \RINGONE)$ is a set $\RING$ with closed binary operations $\RINGPLUS$ and $\RINGPROD$, the additive identity $\RINGZERO$, and the multiplicative identity $\RINGONE$ satisfying the axioms ($\forall a,b,c\in\RING$):
\begin{enumerate}
    \item $a \RINGPLUS b = b\RINGPLUS a$.
    \item $(a \RINGPLUS b)\RINGPLUS c = a \RINGPLUS (b \RINGPLUS c)$.
    \item $\RINGZERO \RINGPLUS a = a \RINGPLUS \RINGZERO = a$.
    \item $\exists -a \in \RING: a \RINGPLUS (-a) = (-a) \RINGPLUS a = \RINGZERO$.
    \item $(a \RINGPROD b) \RINGPROD c = a \RINGPROD (b \RINGPROD c)$.
    \item $a \RINGPROD \RINGONE = \RINGONE * a = a$.
    \item $a \RINGPROD (b \RINGPLUS c) = a \RINGPROD b \RINGPLUS a \RINGPROD c$ and $(a \RINGPLUS b) \RINGPROD c = a \RINGPROD c \RINGPLUS b \RINGPROD c$.
\end{enumerate}
A semiring $(\RING, \RINGPLUS, \RINGPROD, \RINGZERO, \RINGONE)$ satisfies all of the above properties except the additive inverse property (4) and adds the axiom $\RINGZERO \RINGPROD a = a \RINGPROD \RINGZERO = \RINGZERO$.
A (semi)ring for which $a \RINGPROD b = b \RINGPROD a$ is commutative. Axiom (7) is the distributivity law.}, such as work on $k$-relations over provenance semirings~\cite{PROVSEMIRING:PODS:2017}, generalised multiset relations~\cite{DBRING:PODS:2010}, and factors over semirings~\cite{faq}.

The following properties make the rings particularly effective for computing and maintaining aggregates.	

\medskip

{\noindent\bf Distributivity law.} This powerful law underlies our work on factorised databases~\cite{OlZa15,OS:SIGREC:2016}. In particular, it allows to factor out data blocks common to several tuples, represent them once and compute over them once. Using factorisation, relations can be represented more succinctly as directed acyclic graphs. For instance,  we can trade the explicit representation of the Cartesian product of two sets $R=\cup_{i\in[n]} r_i$ and $S=\cup_{j\in[m]} s_j$ for a symbolic representation:
\begin{align*}
	\cup_{i\in[n], j\in[m]} (r_i \times s_j) = \big(\cup_{i\in[n]} r_i\big)\times\big(\cup_{j\in[m]} s_j\big)
\end{align*}
Whereas the left expression has $2nm$ values, the right expression only has $n+m$ values. This can be generalised to joins, which are by definition unions of Cartesian products, and also to queries with select, project, join, group-by aggregates, and order-by clauses. A join query may admit several factorisations and a question is which ones are optimal, i.e., there is no factorisation whose size is smaller in worst-case. Can such factorisations can be computed optimally, i.e., in time proportional to their sizes? These questions have been answered in the affirmative~\cite{OlZa15}. Factorisation can lower the computational complexity of jo\-ins~\cite{FDB:PVLDB:2012}, aggregates~\cite{BKOZ13}, and  machine learning~\cite{SOC:SIGMOD:16,OS:SIGREC:2016,ANNOS:PODS:2018,faqai}. Section~\ref{sec:case1} exemplifies factorised computation for joins and aggregates.

As a representation formalism, factorised databases are a special class of multi-valued ordered decision diagrams that have been studied in knowledge compilation, a field in AI literature that aims at exploiting structure in (worst-case hard) problems to solve them efficiently in practice~\cite{Darwiche:PODS:2020}. The framework of Functional Aggregate Queries~\cite{faq} generalises factorised databases to semirings beyond sum-product and shows that many problems across Computer Science can benefit from factorised computation. 
LMFAO~\cite{lmfao,LMFAO:PVLDB:2020}, F-IVM~\cite{Nikolic:FIVM:2018, FIVM:SIGMOD:2020}, and IVM$^{\epsilon}$~\cite{IVMe:PODS:2020,IVMe:TODS:2020} employ factorised query computation and maintenance. These systems factorise the query into a hierarchy of increasingly simpler views, which are maintained bottom-up under data updates.

\medskip

{\noindent\bf Additive inverse.}  The additive inverse of rings allows to treat uniformly data updates (inserts and deletes). This would require relations to map tuples to their multiplicities, i.e., using the ring of integers. Then, an insert/delete of a tuple $r$ into/from a relation $R$ is modelled as the addition of the mapping $\{r\mapsto 1\}$ and respectively $\{r\mapsto -1\}$ to the map $R$. If the key $r$ is already present in $R$, then we sum up their multiplicities so that each key appears once in the map $R$. Tuples with multiplicity zero are not represented.

This uniform treatment of updates simplifies the maintenance procedure for queries~\cite{Koch:DBToaster:2014,IVMe:ICDT:2019} and models~\cite{Nikolic:FIVM:2018,FIVM:SIGMOD:2020}.

\medskip

{\noindent\bf Sum-product abstraction.} The sum-product abstraction in rings allows to use the same processing (computing and maintaining) mechanism for seemingly disparate tasks, such as database queries, covariance matrices, inference in probabilistic graphical models, and matrix chain multiplication~\cite{faq,Nikolic:FIVM:2018}. Section~\ref{sec:case2} exemplifies a ring designed for covariance matrices. The efficient maintenance of covariance matrices makes it possible to keep models fresh under high-throughput data changes~\cite{Nikolic:FIVM:2018}. A recent tutorial overviews advances in incremental view maintenance~\cite{Iman:CIKM:2019}.

\subsection{Combinatorial structure}

The combinatorial structure prevalent in relational data is captured by notions such as the width measure of the query and the degree of a data value. If a feature extraction query has width $w$, then its data complexity is $\tilde{O}(N^w)$ for a database of size $N$, where $\tilde{O}$ hides logarithmic factors in $N$.

\medskip

{\noindent\bf Width measures.}  Various increasingly more refined and smaller width measures have been proposed recently: 
\begin{itemize}
\item The fractional edge cover number~\cite{GM06} captures the asymptotic size of the results for join queries~\cite{GM06,AGM08} and the time to compute them~\cite{NPRR12,skew,LFTJ}.

\item The fractional hypertree width~\cite{Marx:FHTW:2010} is defined using the fractional edge cover number of the bags in the hypertree decompositions of the query. 

\item The factorisation width~\cite{OlZa15,OS:SIGREC:2016} generalises the fractional hypertree width to non-Boolean  queries. This width defines the size of factorised query results.

\item The FAQ-width~\cite{faq} generalises the factorisation width to functional aggregate queries over several semirings.

\item The submodular width~\cite{SUBW:JACM:2013,panda17} is the smallest known width for Boolean conjunctive queries.

\item The relaxed versions~\cite{faqai} of FAQ-width and submodular width capture the complexity of queries with additive inequalities discussed in Section~\ref{sec:nonpolynomial-loss}.
\end{itemize}

This theoretical development led to a number of algorithms that attain the above width measures: LFTJ~\cite{LFTJ}, F~\cite{SOC:SIGMOD:16,OS:SIGREC:2016}, InsideOut~\cite{faq}, EmptyHeaded~\cite{EmptyHeaded:TODS:2017}, PANDA~\cite{panda17}, and \#PANDA~\cite{faqai}. Our prototypes F~\cite{SOC:SIGMOD:16} and LMFAO~\cite{lmfao} for learning over relational data use an underlying query engine that achieves the factorisation width.

\medskip

{\noindent\bf Data degree.} The degree information captures the number of occurrences of a data value in the input database~\cite{skew}. Several existing query processing algorithms adapt their execution strategy depending on the degree of data values, with different strategies for heavy and light values, where a value is heavy/light if its degree is above/below a certain chosen threshold. This adaptive processing has been used by  worst-case optimal join algorithms~\cite{NPRR12}, worst-case optimal incremental maintenance of triangle queries~\cite{IVMe:ICDT:2019,IVMe:TODS:2020}, and to achieve the submodular width for Boolean conjunctive queries~\cite{panda17}. The complexity of query evaluation can reduce dramatically in the presence of bounded degrees or even more refined notions of data sparsity, such as nowhere denseness~\cite{NOWHEREDENSE:JACM:2017}. This is the case, for instance, for computing queries with negated relations of bounded degree~\cite{Khamis:ICDT:2019}.

\medskip

{\noindent\bf Functional dependencies.} A special form of bounded degree is given by functional dependencies. They can be used to lower the learning runtime for ridge polynomial regression models and factorisation machines. Instead of learning a given model, we can instead learn a reparameterised model with fewer parameters and then map it back to the original model~\cite{ANNOS:PODS:2018,ANNOS:TODS:2020}. Take for instance a model whose parameters include a parameter $\theta_{\textsf{city}}$ for each $\textsf{city}$ and $\theta_{\textsf{country}}$ for each $\textsf{country}$ in the input data, and assume the functional dependency $\textsf{city}\rightarrow\textsf{country}$ holds in the database. We can replace each parameter pair $(\theta_{\textsf{city}},\theta_{\textsf{country}})$ with a new parameter $\theta_{(\textsf{city},\textsf{country})}$. The relationship between the original and new parameters can be expressed in closed form for ridge polynomial regression and factorisation machines, hence $(\theta_{\textsf{city}},\theta_{\textsf{country}})$ can be recovered from $\theta_{(\textsf{city},\textsf{country})}$.

\subsection{Statistical and geometric structure}

An alternative approach to computing over the entire input data is to compute over data samples. There is a solid body of work in sampling for machine learning~\cite{murphy2013}. In our database setting, the major challenge is that we would need to sample through the feature extraction query. The most notable advances in the database literature is on sampling through selection conditions and joins, e.g., the ripple joins~\cite{Haas:RippleJoin:SIGMOD:1999} and the wander joins~\cite{Li:WanderJoin:TODS:2019}. However, the feature extraction query may also compute derived aggregate features, which means that sampling through aggregates would be also needed. More recent work considers sampling for specific classes of machine learning models~\cite{Park:SampleML:SIGMOD:2019}. This is a challenging and interesting subject to future work.

Sampling is employed whenever the input database is too large to be processed within a given time budget. It may nevertheless lead to approximation of both steps in the end-to-end learning task, from the computation of the feature extraction query to the subsequent optimization task that yields the desired model.
Work in this space quantifies the loss in accuracy of the obtained model due to sampling~\cite{murphy2013}.

Geometric structure is relevant whenever we use distance measures. Clustering algorithms can exploit such measures, e.g., the optimal transport distance between two probability measures, and inequalities between them, e.g., the triangle inequality. One example is Relational $k$-means~\cite{Rkmeans:AISTATS:2020}, which achieves constant-factor approximations of the $k$-means objective by clustering over a small coreset instead of the full result of the feature extraction query.

\section{Lowering the constant factors}
\label{sec:constants}

\begin{quote}
	{\em The database researcher continuously refines a toolbox of clever system tricks, including:  specialisation for workload, data, and hardware; observing the memory hierarchy and blocking operations; distribution and parallelisation. These tools can bring otherwise computationally hard problems into the realm of the possible.} 
\end{quote}

\begin{figure}
	\centering
  \includegraphics[scale=.875]{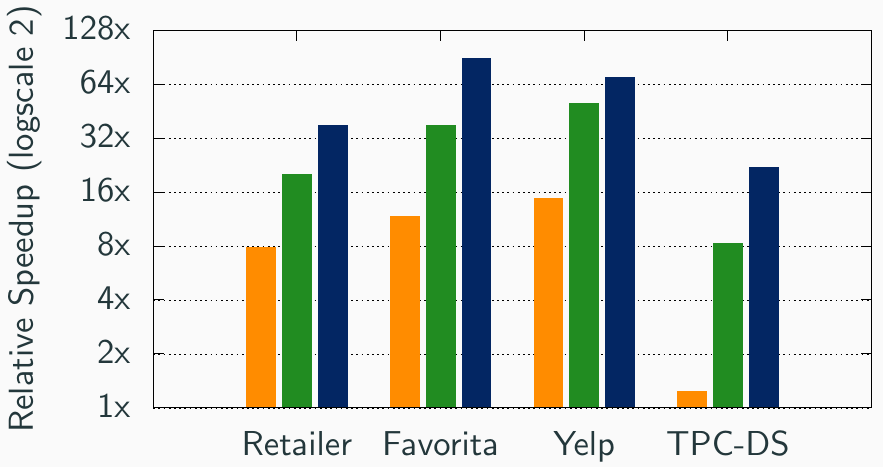}
\caption{Speedup of code optimisations in LMFAO for computing the covariance matrix relative to baseline implementation. The optimisations are added in this order: \textcolor{darkorange}{specialisation}, \textcolor{forestgreen}{sharing},  \textcolor{oxfordblue}{parallelization} (AWS d2.xlarge, 4 vCPUs, 32GB).}
\label{fig:lmfao-breakdown}
\end{figure}

While asymptotic runtime improvements are desirable, their successful implementation in practical cases is not a given. They also aim at upper bounding the worse-case scenario, which may not happen in practice. Worst-case optimal join algorithms~\cite{LFTJ,LB:SIGMOD:2015,LB:Experiments:2015} and factorised computation and maintenance of models~\cite{lmfao,Nikolic:FIVM:2018} have been validated in practice. However, the jury is still out on adaptive query strategies~\cite{IVMe:ICDT:2019,IVMe:PODS:2020} and the submodular width~\cite{panda17}.

At least of equal relevance are systems aspects of theoreti\-cally-efficient algorithms. Our prototyping effort considered code specialisation, sha\-red scans, and parallelisation, which contribute to LMFAO's runtime performance as shown in Figure~\ref{fig:lmfao-breakdown}. The baseline for this comparison (1x speedup) is AC/DC~\cite{ANNOS:DEEM:18}, the precursor of LMFAO that has none of these optimisations. We next mention aspects of lowering the constants in our LMFAO and IFAQ systems. 

\medskip

{\noindent\bf Workload compilation.} LMFAO generates code specific to the query and database schema~\cite{Neumann:PVLDB:11,legobase_tods,dblablb}, and also specific to the model. 

For a more systematic study of the code optimisation and generation effort in LMFAO, we recently initiated a new project called IFAQ (Iterative Functional Aggregate Queries)~\cite{IFAQ:CGO:2020}. IFAQ is a framework that allows to specify in a unified domain-specific language DB+ML workloads and provides a unified optimisation and compilation approach to programs in this language. Example of such programs include: gradient descent optimisation for learning a variety of models over relational data, the CART algorithm for decision trees, model selection, and cross validation. Section~\ref{sec:case3} exemplifies IFAQ transformations for learning a linear regression model over a join using gradient descent.

\medskip

{\noindent\bf Sharing computation.} Both LMFAO and IFAQ exhibit sharing across the batch of aggregates at different stages. We next sketch two common sharing scenarios. 

Consider a join tree of the feature extraction query\footnote{If the query is cyclic, it is first partially evaluated to an acyclic query, e.g., by materialising the bags of any of its hypertree decomposition~\cite{GottlobTD}.}. We decompose each aggregate in the batch following a top-down traversal of this tree as follows. We assign at a node in the tree a restriction of the aggregate that only refers to the attributes  within the subtree rooted at that node. This means that we assign the entire aggregate at the root. It also means that if the subtree rooted at a node has no attributes in the aggregate, then we assign a count aggregate at the node. The reason is that this restriction of the aggregate can be answered at each node in the tree using the partial aggregates computed at its children, if any. At each node, we then consolidate the synthesised aggregates into views. For the batches discussed in Section~\ref{sec:aggregates}, it is often the case that several aggregates have the same partial aggregates at a node in the tree, so we only compute these partial common aggregates once and use them several times.

Furthermore, to compute the views at the same node, we can share the scan of the relation at the node. To increase sharing and minimise the sizes of the intermediate views, LMFAO may decompose different aggregates starting at different root nodes in the join tree.

\medskip

{\noindent\bf Parallelisation.} LMFAO exploits multi-core CPU architectures. It supports task parallelism by computing in parallel groups of views that do not depend on each other. It supports domain parallelism by deploying threads to compute view groups on disjoint parts of a relation.

\nop{
{\noindent\bf Parallelisation.} LMFAO exploits multi-core CPU architectures. Task parallelism is enabled by grouping the views at the same node in the join tree and then creating a (acyclic) dependency graph between the groups, where one group depends on the other if the output of the latter is needed as the input to the former. LMFAO assigns independent groups to different threads. Domain parallelism is performed for the computation of the views within each group by having different threads compute the views over disjoint parts of the relation  at a node in the join tree.
}

\begin{figure*}
\begin{scriptsize}
\[
\begin{tabular}{@{~}rrr@{~}}
\multicolumn{3}{c}{\textrm{Orders}}\\\toprule
 customer & day & dish \\\midrule
 Elise & Monday & burger \\
 Elise & Friday & burger \\
 Steve & Friday & hotdog \\
 Joe & Friday & hotdog \\\bottomrule
\ \\
\ \\
\end{tabular}
\quad
\begin{tabular}{@{~}rr@{~}}
\multicolumn{2}{c}{\textrm{Dish}}\\\toprule
dish & item \\\midrule
 burger & patty \\
 burger & onion \\
 burger & bun \\
 hotdog & bun \\
 hotdog & onion \\
 hotdog & sausage \\\bottomrule
\end{tabular}
\quad
\begin{tabular}{@{~}rrr@{~}}
\multicolumn{2}{c}{\textrm{Items}}\\\toprule
 item & price\\\midrule
 patty & 6 \\
 onion & 2 \\
 bun & 2 \\
 sausage & 4 \\\bottomrule
\ \\
\ \\
\end{tabular}
\quad
\begin{tabular}{@{}r@{ $\times$ }r@{ $\times$ }r@{ $\times$ }r@{ $\times$ }r@{ $\cup$ }r@{}}
\multicolumn{5}{c}{Natural join of Orders, Dish, and Items}\\\toprule
customer  & day & dish & item & price \\\midrule
Elise     & {\color{goodgreen}Monday} &  {\color{teal}burger}  & {\color{red}\mbox{patty}} & 6  \\
Elise     & {\color{goodgreen}Monday} &  {\color{teal}burger}  & {\color{amber}\mbox{onion}} & 2  \\
Elise     & {\color{goodgreen}Monday} &  {\color{teal}burger}  & {\color{burntorange}\mbox{bun}} & 2  \\
Elise    & {\color{blue}Friday} &  {\color{teal}burger}  & {\color{red}\mbox{patty}} & 6  \\
Elise     & {\color{blue}Friday} &  {\color{teal}burger}  & {\color{amber}\mbox{onion}} & 2  \\
Elise     & {\color{blue}Friday} &  {\color{teal}burger}  & {\color{burntorange}\mbox{bun}} & 2  \\\bottomrule
\end{tabular}
\]
\end{scriptsize}

\caption{Example database used in Section~\ref{sec:case1} with relations Orders, Dish, and Items, and a fragment of their natural join expressed as a relational algebra expression with union and Cartesian product.}
\label{fig:db}
\end{figure*}

\begin{figure*}
\centering
\begin{tikzpicture}[xscale=0.55, yscale=0.55]
\begin{scriptsize}

\node at (0, 0) (u1) {$\cup$};

\node at (-5, -1) (d1) {$\color{teal}burger$} edge[-] (u1);
\node at (4, -1) (d2) {$\color{applegreen}hot dog$} edge[-] (u1);

\node at (-5, -2) (x1) {$\times$} edge[-] (d1);
\node at (4, -2) (x2) {$\times$} edge[-] (d2);

\node at (1.8, -3) (u1) {$\cup$} edge[-] (x2);

\node at (3.5, -4) (i1) {$sausage$} edge[-] (u1);
\node at (0.5, -4) (i2) {$\color{burntorange}bun$} edge[-] (u1);
\node at (1.8, -4) (i3) {$\color{amber}onion$} edge[-] (u1);

\node at (3.5, -5) (x3) {$\times$} edge[-] (i1);
\node at (0.5, -5) (x4) {$\times$} edge[-] (i2);
\node at (1.8, -5) (x5) {$\times$} edge[-] (i3);

\node at (3.5, -6) (u2) {$\cup$} edge[-] (x3);

\node at (3.5, -7) (p1) {$4$} edge[-] (u2);

\node at (6, -3) (u5) {$\cup$} edge[-] (x2);
\node at (6, -4) (D1) {$\color{blue}Friday$} edge[-] (u5);
\node at (6, -5) (x6) {$\times$} edge[-] (D1);
\node at (6, -6) (u6) {$\cup$} edge[-] (x6);
\node at (4.7, -7) (N1) {$Joe$} edge[-] (u6);
\node at (7, -7) (N2) {$Steve$} edge[-] (u6);

\node at (-2.5, -3) (u7) {$\cup$} edge[-] (x1);
\node at (-4, -4) (i5) {$\color{red}patty$} edge[-] (u7);
\node at (-2.5, -4) (i6) {$\color{burntorange}bun$} edge[-] (u7);
\node at (-1, -4) (i7) {$\color{amber}onion$} edge[-] (u7);

\node at (-4, -5) (x7) {$\times$} edge[-] (i5);
\node at (-2.5, -5) (x8) {$\times$} edge[-] (i6);
\node at (-1, -5) (x9) {$\times$} edge[-] (i7);

\node at (-4, -6) (u8) {$\cup$} edge[-] (x7);
\node at (-2.5, -6) (u9) {$\cup$} edge[-] (x8);
\node at (-1, -6) (u10) {$\cup$} edge[-] (x9);

\node at (-4, -7) (p1) {$\color{red}6$} edge[-] (u8);
\node at (-2.5, -7) (p2) {$\color{burntorange}2$} edge[-] (u9);
\node at (-1, -7) (p3) {$\color{amber}2$} edge[-] (u10);

\node at (-7, -3) (u11) {$\cup$} edge[-] (x1);
\node at (-6, -4) (D2) {$\color{blue}Friday$} edge[-] (u11);
\node at (-6, -5) (x10) {$\times$} edge[-] (D2);
\node at (-6, -6) (u12) {$\cup$} edge[-] (x10);
\node at (-6, -7) (p1) {$Elise$} edge[-] (u12);

\node at (-8, -4) (D2) {$\color{goodgreen}Monday$} edge[-] (u11);
\node at (-8, -5) (x10) {$\times$} edge[-] (D2);
\node at (-8, -6) (u12) {$\cup$} edge[-] (x10);
\node at (-8, -7) (p1) {$Elise$} edge[-] (u12);

\draw[dashed] (x4) -- (u9);
\draw[dashed] (x5) -- (u10);

\node at (-12, -1) (dt0) {$dish$};
\node at (-13, -1)       {$\emptyset$};

\node at (-13, -4) (dt1) {$day$} edge[-] (dt0);
\node at (-13.75, -3.5)       {$\{dish\}$};
\node at (-11, -4) (dt3) {$item$} edge[-] (dt0);
\node at (-10.25, -3.5)        {$\{dish\}$};

\node at (-13, -7) (dt2) {$customer$} edge[-] (dt1);
\node at (-13.75, -6)       {$\{dish,$};
\node at (-13.75, -6.5)       {$day\}$};
\node at (-11, -7) (dt4) {$price$} edge[-] (dt3);
\node at (-10.25, -6.5)        {$\{item\}$};
\end{scriptsize}

\end{tikzpicture}

\caption{Factorised representation of the join result in Figure~\ref{fig:db}. Left: Order on the query variables, where each variable is adorned with its ancestors on which it depends. Right: Factorisation of the natural join from Figure~\ref{fig:db} modelled on the variable order.}
\label{fig:factorised-join}
\end{figure*}
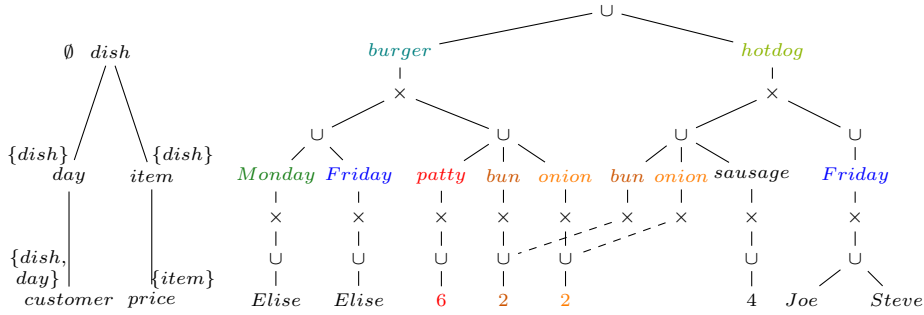

\begin{figure*}
\begin{tikzpicture}[xscale=0.55, yscale=0.55]
\begin{scriptsize}

\node at (0, 0) (u1) {$+$};

\node at (-5, -1) (d1) {$\color{teal} 1$} edge[-] (u1);
\node at (4, -1) (d2) {$\color{applegreen} 1$} edge[-] (u1);

\node at (-5, -2) (x1) {$*$} edge[-] (d1);
\node at (4, -2) (x2) {$*$} edge[-] (d2);

\node at (1.8, -3) (u1) {$+$} edge[-] (x2);
\node at (3.5, -4) (i1) {$1$} edge[-] (u1);
\node at (0.5, -4) (i2) {$\color{burntorange} 1$} edge[-] (u1);
\node at (1.8, -4) (i3) {$\color{amber}1$} edge[-] (u1);

\node at (3.5, -5) (x3) {$*$} edge[-] (i1);
\node at (0.5, -5) (x4) {$*$} edge[-] (i2);
\node at (1.8, -5) (x5) {$*$} edge[-] (i3);

\node at (3.5, -6) (u2) {$+$} edge[-] (x3);

\node at (3.5, -7) (p1) {$1$} edge[-] (u2);

\node at (6, -3) (u5) {$+$} edge[-] (x2);
\node at (6, -4) (D1) {$\color{blue} 1$} edge[-] (u5);
\node at (6, -5) (x6) {$*$} edge[-] (D1);
\node at (6, -6) (u6) {$+$} edge[-] (x6);
\node at (4.7, -7) (N1) {$ 1$} edge[-] (u6);
\node at (7, -7) (N2) {$ 1$} edge[-] (u6);

\node at (-2.5, -3) (u7) {$+$} edge[-] (x1);
\node at (-4, -4) (i5) {$\color{red} 1$} edge[-] (u7);
\node at (-2.5, -4) (i6) {$\color{burntorange} 1$} edge[-] (u7);
\node at (-1, -4) (i7) {$\color{amber} 1$} edge[-] (u7);

\node at (-4, -5) (x7) {$*$} edge[-] (i5);
\node at (-2.5, -5) (x8) {$*$} edge[-] (i6);
\node at (-1, -5) (x9) {$*$} edge[-] (i7);

\node at (-4, -6) (u8) {$+$} edge[-] (x7);
\node at (-2.5, -6) (u9) {$+$} edge[-] (x8);
\node at (-1, -6) (u10) {$+$} edge[-] (x9);

\node at (-4, -7) (p1) {$\color{red} 1$} edge[-] (u8);
\node at (-2.5, -7) (p2) {$\color{burntorange} 1$} edge[-] (u9);
\node at (-1, -7) (p3) {$\color{amber}1$} edge[-] (u10);

\node at (-7, -3) (u11) {$+$} edge[-] (x1);
\node at (-6, -4) (D2) {$\color{blue} 1$} edge[-] (u11);
\node at (-6, -5) (x10) {$*$} edge[-] (D2);
\node at (-6, -6) (u12) {$+$} edge[-] (x10);
\node at (-6, -7) (p1) {$1$} edge[-] (u12);

\node at (-8, -4) (D2) {$\color{goodgreen} 1$} edge[-] (u11);
\node at (-8, -5) (x10) {$*$} edge[-] (D2);
\node at (-8, -6) (u12) {$+$} edge[-] (x10);
\node at (-8, -7) (p1) {$1$} edge[-] (u12);

\draw[dashed] (x4) -- (u9);
\draw[dashed] (x5) -- (u10);

\node[draw,rectangle,scale=0.9,red] at (0, -0.7) (cnt-x) {$12$};

\node[draw,rectangle,scale=0.9,red] at (3.3, -1.8) (cnt-x) {$6$};
\node[draw,rectangle,scale=0.9,red] at (-5.7, -1.8) (cnt-x) {$6$};

\node[draw,rectangle,scale=0.9,red] at (-7.7, -2.8) (cnt-x) {$2$};

\node[draw,rectangle,scale=0.9,red] at (-1.9, -2.8) (cnt-x) {$3$};

\node[draw,rectangle,scale=0.9,red] at (-7.5, -5.5) (cnt-x) {$1$};
\node[draw,rectangle,scale=0.9,red] at (-5.5, -5.5) (cnt-x) {$1$};
\node[draw,rectangle,scale=0.9,red] at (-3.5, -5.5) (cnt-x) {$1$};
\node[draw,rectangle,scale=0.9,red] at (-1.9, -6.3) (cnt-x) {$1$};
\node[draw,rectangle,scale=0.9,red] at (-0.4, -6.3) (cnt-x) {$1$};

\node[draw,rectangle,scale=0.9,red] at (1.2, -2.8) (cnt-x) {$3$};
\node[draw,rectangle,scale=0.9,red] at (6.6, -2.8) (cnt-x) {$2$};


\node[draw,rectangle,scale=0.9,red] at (4, -5.5) (cnt-x) {$1$};
\node[draw,rectangle,scale=0.9,red] at (6.5, -5.5) (cnt-x) {$2$};

\end{scriptsize}
\end{tikzpicture}%
\hspace*{1em}%
\begin{tikzpicture}[xscale=0.55, yscale=0.55]
\begin{scriptsize}

\node at (0, 0) (u1) {$+$};

\node at (-5, -1) (d1) {$f\,({\color{teal} burger}$)} edge[-] (u1);
\node at (4, -1) (d2) {$f\,({\color{applegreen} hotdog}$)} edge[-] (u1);

\node at (-5, -2) (x1) {$*$} edge[-] (d1);
\node at (4, -2) (x2) {$*$} edge[-] (d2);

\node at (1.8, -3) (u1) {$+$} edge[-] (x2);
\node at (3.5, -4) (i1) {$1$} edge[-] (u1);
\node at (0.5, -4) (i2) {$\color{burntorange} 1$} edge[-] (u1);
\node at (1.8, -4) (i3) {$\color{amber}1$} edge[-] (u1);

\node at (3.5, -5) (x3) {$*$} edge[-] (i1);
\node at (0.5, -5) (x4) {$*$} edge[-] (i2);
\node at (1.8, -5) (x5) {$*$} edge[-] (i3);

\node at (3.5, -6) (u2) {$+$} edge[-] (x3);

\node at (3.5, -7) (p1) {$4$} edge[-] (u2);

\node at (6, -3) (u5) {$+$} edge[-] (x2);
\node at (6, -4) (D1) {$\color{blue} 1$} edge[-] (u5);
\node at (6, -5) (x6) {$*$} edge[-] (D1);
\node at (6, -6) (u6) {$+$} edge[-] (x6);
\node at (4.7, -7) (N1) {$ 1$} edge[-] (u6);
\node at (7, -7) (N2) {$ 1$} edge[-] (u6);

\node at (-2.5, -3) (u7) {$+$} edge[-] (x1);
\node at (-4, -4) (i5) {$\color{red} 1$} edge[-] (u7);
\node at (-2.5, -4) (i6) {$\color{burntorange} 1$} edge[-] (u7);
\node at (-1, -4) (i7) {$\color{amber} 1$} edge[-] (u7);

\node at (-4, -5) (x7) {$*$} edge[-] (i5);
\node at (-2.5, -5) (x8) {$*$} edge[-] (i6);
\node at (-1, -5) (x9) {$*$} edge[-] (i7);

\node at (-4, -6) (u8) {$+$} edge[-] (x7);
\node at (-2.5, -6) (u9) {$+$} edge[-] (x8);
\node at (-1, -6) (u10) {$+$} edge[-] (x9);

\node at (-4, -7) (p1) {$\color{red}6$} edge[-] (u8);
\node at (-2.5, -7) (p2) {${\color{burntorange}2}$} edge[-] (u9);
\node at (-1, -7) (p3) {$\color{amber}{2}$} edge[-] (u10);

\node at (-7, -3) (u11) {$+$} edge[-] (x1);
\node at (-6, -4) (D2) {$\color{blue} 1$} edge[-] (u11);
\node at (-6, -5) (x10) {$*$} edge[-] (D2);
\node at (-6, -6) (u12) {$+$} edge[-] (x10);
\node at (-6, -7) (p1) {$1$} edge[-] (u12);

\node at (-8, -4) (D2) {$\color{goodgreen} 1$} edge[-] (u11);
\node at (-8, -5) (x10) {$*$} edge[-] (D2);
\node at (-8, -6) (u12) {$+$} edge[-] (x10);
\node at (-8, -7) (p1) {$1$} edge[-] (u12);

\draw[dashed] (x4) -- (u9);
\draw[dashed] (x5) -- (u10);

\node[draw,rectangle,scale=0.9,red] at (0, 0.7) (cnt-x) {$20{\color{black} * f\,({\color{teal} burger})} {\color{black}+} 16  {\color{black} * f\,({\color{applegreen} hotdog})}$};

\node[draw,rectangle,scale=0.9,red] at (3.3, -1.8) (cnt-x) {$16$};
\node[draw,rectangle,scale=0.9,red] at (-5.7, -1.8) (cnt-x) {$20$};

\node[draw,rectangle,scale=0.9,red] at (-7.7, -2.8) (cnt-x) {$2$};

\node[draw,rectangle,scale=0.9,red] at (-1.7, -2.7) (cnt-x) {$10$};

\node[draw, rectangle,scale=0.9,red] at (-7.5, -5.5) (cnt-x) {$1$};
\node[draw, rectangle,scale=0.9,red] at (-5.5, -5.5) (cnt-x) {$1$};

\node[draw,rectangle,scale=0.9,red] at (-3.5, -5.5) (cnt-x) {$6$};
\node[draw,rectangle,scale=0.9,red] at (-2, -6.4) (cnt-x) {$2$};
\node[draw,rectangle,scale=0.9,red] at (-0.5, -6.4) (cnt-x) {$2$};

\node[draw, rectangle,scale=0.9,red] at (1.2, -2.7) (cnt-x) {$8$};
\node[draw, rectangle,scale=0.9,red] at (6.2, -2.4) (cnt-x) {$2$};


\node[draw, rectangle,scale=0.9,red] at (4.2, -5.5) (cnt-x) {$4$};
\node[draw, rectangle,scale=0.9,red] at (6.5, -5.5) (cnt-x) {$2$};

\end{scriptsize}
\end{tikzpicture}

\caption{Aggregate computation computed in one pass over the factorised join in Figure~\ref{fig:factorised-join}. Left: \texttt{SUM(1)}. We map all values to 1, $\cup$ to $+$ and $\times$ to \texttt{*}. Right: \texttt{SUM(dish * price)}. We assume there is a function $f$ that turns \texttt{dish} into numbers (e.g., one-hot encoding) or we group by \texttt{dish}: \texttt{SUM(price) GROUP BY dish}. We map all values except for \texttt{dish} and \texttt{price} to 1, $\cup$ to $+$ and $\times$ to \texttt{*}.}
\label{fig:factorising-aggregates}
\end{figure*}
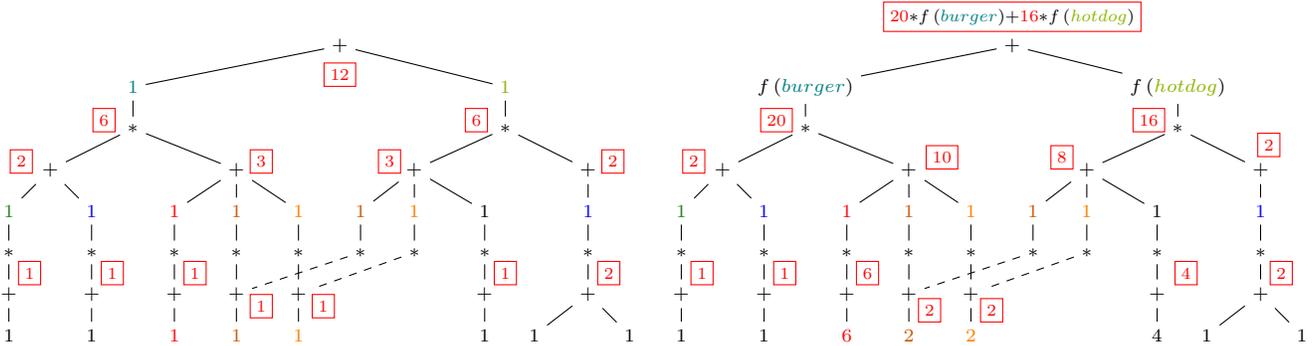

\begin{figure*}
\begin{tikzpicture}[xscale=0.55, yscale=0.55]
\begin{scriptsize}

\node at (-5, -1) (d1) {$\color{teal}burger$};

\node at (-5, -2) (x1) {$\times$} edge[-] (d1);

\node at (-1.5, -3) (u7) {$\cup$} edge[-] (x1);
\node at (-3, -4) (i5) {$\color{red}patty$} edge[-] (u7);
\node at (-1.5, -4) (i6) {$\color{burntorange}bun$} edge[-] (u7);
\node at (.5, -4) (i7) {$\color{amber}onion$} edge[-] (u7);

\node at (-3, -5) (x7) {$\times$} edge[-] (i5);
\node at (-1.5, -5) (x8) {$\times$} edge[-] (i6);
\node at (.5, -5) (x9) {$\times$} edge[-] (i7);

\node at (-3, -6) (u8) {$\cup$} edge[-] (x7);
\node at (-1.5, -6) (u9) {$\cup$} edge[-] (x8);
\node at (.5, -6) (u10) {$\cup$} edge[-] (x9);

\node at (-3, -7) (p1) {$\color{red}6$} edge[-] (u8);
\node at (-1.5, -7) (p2) {$\color{burntorange}2$} edge[-] (u9);
\node at (.5, -7) (p3) {$\color{amber}2$} edge[-] (u10);

\node at (-8, -3) (u11) {$\cup$} edge[-] (x1);
\node at (-6.5, -4) (D2) {$\color{blue}Friday$} edge[-] (u11);
\node at (-6.5, -5) (x10) {$\times$} edge[-] (D2);
\node at (-6.5, -6) (u12) {$\cup$} edge[-] (x10);
\node at (-6.5, -7) (p1) {$Elise$} edge[-] (u12);

\node at (-9.5, -4) (D2) {$\color{goodgreen}Monday$} edge[-] (u11);
\node at (-9.5, -5) (x10) {$\times$} edge[-] (D2);
\node at (-9.5, -6) (u12) {$\cup$} edge[-] (x10);
\node at (-9.5, -7) (p1) {$Elise$} edge[-] (u12);

\end{scriptsize}
\end{tikzpicture}%
\hspace*{1em}%
\begin{tikzpicture}[xscale=0.55, yscale=0.55]
\begin{scriptsize}

\node at (-5, -1) (d1) {$(1,0,f({\color{teal}burger}))$};

\node at (-5, -2) (x1) {$*$} edge[-] (d1);

\node at (0.5, -3) (u7) {$+$} edge[-] (x1);
\node at (-3, -4) (i5) {$(1,0,0)$} edge[-] (u7);
\node at (0.5, -4) (i6) {$(1,0,0)$} edge[-] (u7);
\node at (4, -4) (i7) {$(1,0,0)$} edge[-] (u7);

\node at (-3, -5) (x7) {$*$} edge[-] (i5);
\node at (0.5, -5) (x8) {$*$} edge[-] (i6);
\node at (4, -5) (x9) {$*$} edge[-] (i7);

\node at (-3, -6) (u8) {$+$} edge[-] (x7);
\node at (0.5, -6) (u9) {$+$} edge[-] (x8);
\node at (4, -6) (u10) {$+$} edge[-] (x9);

\node at (-3, -7) (p1) {$(1,{\color{red}6},0)$} edge[-] (u8);
\node at (0.5, -7) (p2) {$(1,{\color{burntorange}2},0)$} edge[-] (u9);
\node at (4, -7) (p3) {$(1,{\color{amber}2},0)$} edge[-] (u10);

\node at (-10, -3) (u11) {$+$} edge[-] (x1);
\node at (-6.5, -4) (D2) {$(1,0,0)$} edge[-] (u11);
\node at (-6.5, -5) (x10) {$*$} edge[-] (D2);
\node at (-6.5, -6) (u12) {$+$} edge[-] (x10);
\node at (-6.5, -7) (p1) {$(1,0,0)$} edge[-] (u12);

\node at (-13.5, -4) (D2) {$(1,0,0)$} edge[-] (u11);
\node at (-13.5, -5) (x10) {$*$} edge[-] (D2);
\node at (-13.5, -6) (u12) {$+$} edge[-] (x10);
\node at (-13.5, -7) (p1) {$(1,0,0)$} edge[-] (u12);

\node[draw,rectangle,scale=0.9,red] at (-12, -2.5) (cnt-x) {$(2,0,0)$};
\node[draw,rectangle,scale=0.9,red] at (2, -2.5) (cnt-x) {$(3,10,0)$};
\node[draw,rectangle,scale=0.9,red] at (-5, -2.8) (cnt-x) {$(2\cdot 3,2\cdot 10,0)$};
\node[draw,rectangle,scale=0.9,red] at (-0.5, -1) (cnt-x) {$(6,20,20\cdot f({\color{teal}burger}))$};

\node[draw,rectangle,scale=0.9,red] at (-8.5, -5) (cnt-x) {$(1,0,0)$};

\node[draw,rectangle,scale=0.9,red] at (-11.5, -5) (cnt-x) {$(1,0,0)$};


\node[draw,rectangle,scale=0.9,red] at (-4.5, -5) (cnt-x) {$(1,6,0)$};
\node[draw,rectangle,scale=0.9,red] at (-1, -5) (cnt-x) {$(1,2,0)$};
\node[draw,rectangle,scale=0.9,red] at (2.5, -5) (cnt-x) {$(1,2,0)$};

\end{scriptsize}

\end{tikzpicture}

\caption{Left: Factorised join. Right: Computing \texttt{SUM(1), SUM(price), SUM(price * dish)} using a ring whose elements are triples of numbers, one per aggregate. The sum (+)  and product (*) operations are now defined over triples. This specialised ring enables shared computation across the aggregates.}
\label{fig:ring-shared-computation}
\end{figure*}
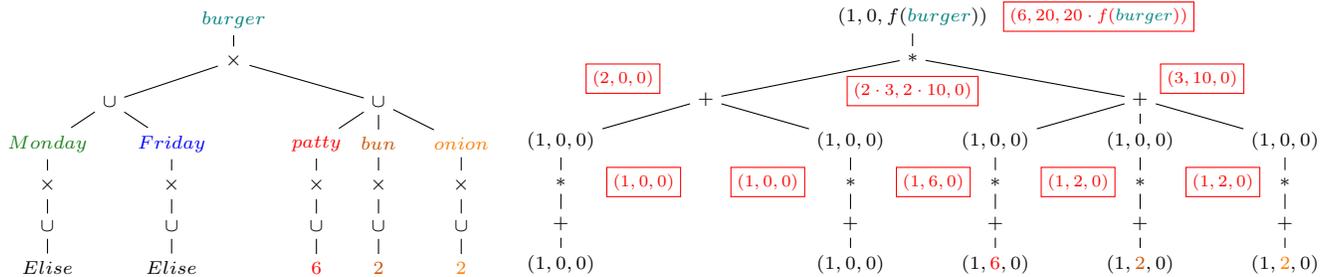

\section{Highlights}
\label{sec:highlights}

We exemplify some of the query computation and maintenance techniques mentioned in Sections~\ref{sec:asymptotics} and \ref{sec:constants}.

\subsection{Factorised computation}
\label{sec:case1}

\begin{quote}
	{\em Factorisation is a normal form for the query result that comes with time and space improvement.} 
\end{quote}

Consider the  relations  and their natural join in Figure~\ref{fig:db}. The relations, including the join, can be interpreted as relational algebra expressions over union and Cartesian product: Each tuple is a product of values and a relation is a union of such tuples. The set of possible relations together with these two operations form a semiring\footnote{Constraint: Only relations with the same schema can be added and relations with disjoint schemas can be multiplied.}.

As highlighted in Figure~\ref{fig:db}, the same value may occur several times in a relation. Using the distributivity and commutativity laws of the semiring, we may factor out occurrences of the same value or even entire expressions so as to achieve a small representation. Figure~\ref{fig:factorised-join} depicts a factorisation of the join. There are several algebraically equivalent factorisations. The one we depict has fewer data values than the input relations. In contrast, the number of values in the non-factorised join can be much larger.

The factorised join is modelled on an order of the variables (or attributes) in the join result. This variable order dictates the compression factor of the factorisation. It exploits the join dependencies in the join result to avoid redundancy in the representation. In this sense, the factorisation can be thought of as a normal form for the join result (in analogy with the normalisation commonly applied to the input relations). To avoid redundancy, the factorisation represents separately conditionally independent information. For instance, the days are independent of the items {\em given} a dish. Indeed, the latter and the former are in different relations, which are joined via dish. We express graphically this conditional independence by having these variables in different branches under dish. Not every conditional independence can be captured by branching. For instance, price is independent of dish given item, yet all three variables are along the same path in the order. To signal this independence, we explicitly adorn each variable in the order by the set of its ancestors on which it depends. For instance, price is adorned with item only and not also with dish.

The structure of the variable order is mirrored in the factorised join. We first group by dish; for each dish, we represent separately the days and customers from the items and their prices. Furthermore, since price is independent from dish given item, we cache the price for a specific item so as to avoid its repetition under several dishes.

By avoiding data repetitions, factorised representations can be much more succinct than their equivalent tabular representation. The variable order can also effectively guide us to compute the factorisation directly from the input relations and in time proportional to its size.

We can compute aggregates of the form given in Section~\ref{sec:aggregates} directly on the factorised join. Moreover, the operators for join and aggregates can be fused so that we do not need to first compute the factorised join. Figure~\ref{fig:factorising-aggregates} shows how to compute a count and a sum-product aggregate used for the covariance matrix. They are computed in one bottom-up traversal of the factorised join using a different semiring. For the count, we use the semiring of the natural numbers and map each value to 1. For the sum-product aggregate, we use the same semiring mapping, with the exception that the price values are kept (since we want to sum over them) while the dish values are mapped to numbers (or become keys, if we were to group by dish).

\subsection{Sum-product abstraction}
\label{sec:case2}

\begin{quote}
	{\em Ring operations can be redefined to capture the shared computation of aggregates in a batch.} 
\end{quote}

We continue the example in Section~\ref{sec:case1} and consider a strategy that shares the computation of several aggregates. This is motivated by the large number of similar aggregates needed to train models, cf.\@ Section~\ref{sec:aggregates}. In particular, we would like to design a ring that captures this shared computation.

Figure~\ref{fig:ring-shared-computation} gives a factorised join (a fragment of the one in Figure~\ref{fig:factorised-join} for simplicity) and a bottom-up run over it annotated with triples of values corresponding to local computations for the aggregates \texttt{SUM(1), SUM(price), SUM(price * dish)}. We can observe that our aggregates are of increasing complexity: the first does not have a variable, the second has one variable, and the third has two variables. Such aggregates are common in covariance matrices.

We proceed with our computation as in Figure~\ref{fig:factorising-aggregates}, yet now with all three aggregates together. There are two instances where computation sharing occurs. At the product node under the root: The count from the left branch is used to scale the sum of prices from the right. At the root node, the second aggregate is used to compute the third aggregate.

A natural question is whether we can generalise this to an entire covariance matrix. For a set of numerical features $x_1,\ldots,x_n$, this matrix is defined by the aggregates \texttt{SUM(1)}, \texttt{SUM}$(x_i)$, and \texttt{SUM}$(x_i* x_j)$ for $i,j\in[n]$. Following our example, we can use their partial computation as follows:
\begin{itemize}
    \item \texttt{SUM(1)} can be reused for all \texttt{SUM}($x_i$) and \texttt{SUM}($x_i * x_j$).
    \item  \texttt{SUM}($x_i$) can be reused for all \texttt{SUM}($x_i * x_j$).
\end{itemize}

The ring $({\cal R},+,*,{\bf 0},{\bf 1})$ over triples of aggregates $(c,{\color{airforceblue}\mv s},{\color{TolDarkPurple}\mv Q})$ captures this shared computation:
\begin{align*}
  (c_1,{\color{airforceblue}\mv s_1},{\color{TolDarkPurple}\mv Q_1}) + (c_2,{\color{airforceblue}\mv s_2},{\color{TolDarkPurple}\mv Q_2}) & = (c_1+c_2,{\color{airforceblue}\mv s_1}+{\color{airforceblue}\mv s_2},{\color{TolDarkPurple}\mv Q_1} + {\color{TolDarkPurple}\mv Q_2})\\
  (c_1,{\color{airforceblue}\mv s_1},{\color{TolDarkPurple}\mv Q_1}) * (c_2,{\color{airforceblue}\mv s_2},{\color{TolDarkPurple}\mv Q_2}) & = (c_1\cdot c_2,c_2\cdot {\color{airforceblue}\mv s_1} + c_1\cdot {\color{airforceblue}\mv s_2}, c_2\cdot {\color{TolDarkPurple}\mv Q_1} +\\
  &\hspace*{1.8em} c_1\cdot {\color{TolDarkPurple}\mv Q_2} +{\color{airforceblue}\mv s_1}{\color{airforceblue}\mv s_2^{T}} + {\color{airforceblue}\mv s_2} {\color{airforceblue}\mv s_1^{T}} )\\
  \mv 0 & = (0,\mv 0_{n\times 1},\mv 0_{n\times n})\\
  \mv 1 & = (1,\mv 0_{n\times 1},\mv 0_{n\times n})
\end{align*}

A ring element consists of a scalar, a vector, and a matrix:
  \vspace*{-1.75em}
  \begin{center}
      \includegraphics[scale=.25]{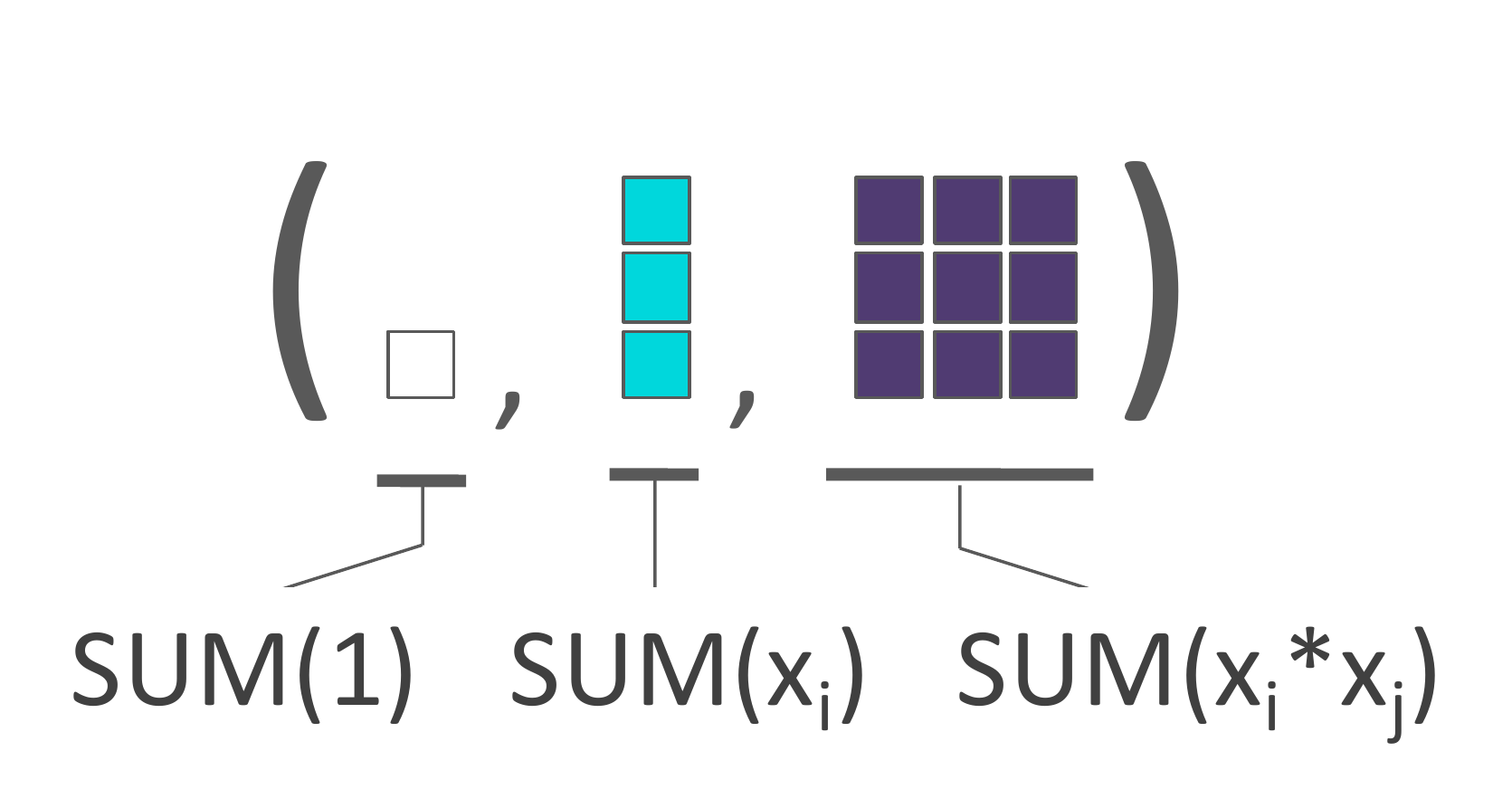}
  \end{center}
  \vspace*{-1.25em}
In the definition of the product operation, the scalars $c_i$ are used to scale the vectors $s_i$ and matrices $Q_i$, while the vectors $s_i$ are used to compute the matrices $Q_i$ ($i\in[2]$). 

Shared computation across the aggregates as emulated by the above ring is partly responsible for the performance benefits shown in Figure~\ref{fig:existing-technology} for LMFAO~\cite{lmfao} and F-IVM~\cite{Nikolic:FIVM:2018}.

\subsection{Multi-stage compilation}
\label{sec:case3}

\begin{figure}
\vspace{0.5cm}
  \begin{tikzpicture}[xscale=0.7, yscale=0.7, transform canvas={scale=0.65}]
    
    \tikzstyle{prog} = [draw, rectangle, align=center,
    inner sep = .1cm, outer sep = .1 cm]
    \tikzstyle{path} = [->, double, thick ,rounded corners=.1cm]

    \node[prog] at (1,0) (aggs) {IFAQ \\ Expr.};

    \node[data,draw=hl_color,color=hl_color] at ($(aggs) + (3,0)$) (sched) {Loop \\ Scheduling};
    \node[data,draw=hl_color,color=hl_color] at ($(sched) + (4,0)$) (factor) {Factorisation};
    \node[data,draw=hl_color,color=hl_color] at ($(factor) + (4.3,0)$) (memo) {Static \\ Memoisation};
    \node[data,draw=hl_color,color=hl_color] at ($(memo) + (3.7,0)$) (motion) {Code \\ Motion};

    \draw[draw=hl_color, rounded corners = .1cm, dashed]
    ($(sched)+(-1.7,1)$) rectangle ($(motion)+(1.4,-1.4)$);
    \node[color=hl_color, anchor=west, scale=1.1] at ($(sched)+(-1.3,-1.1)$)
    {\bf High-Level Optimisations};

    \node[data,draw=sch_color,color=sch_color] at ($(sched) + (-2,-3.2)$) (unroll) {Loop\\ Unrolling};
    \node[data,draw=sch_color,color=sch_color] at ($(unroll) + (3.85,0)$) (static) {Static Field \\ Access};

    \draw[draw=sch_color, rounded corners = .1cm, dashed]
    ($(unroll)+(-1.5,1)$) rectangle ($(static)+(1.8,-1.4)$);
    \node[color=sch_color, anchor=west, scale=1.1] at ($(unroll)+(-1.2,-1.1)$)
    {\bf Schema Specialisation};

    \node[data,draw=agg_color,color=agg_color] at ($(static) + (3.85,0)$) (pushdown) {Aggregate \\ Pushdown};
    \node[data,draw=agg_color,color=agg_color] at ($(pushdown) + (3.85,0)$) (fusion) {Aggregate \\ Fusion};

    \draw[draw=agg_color, rounded corners = .1cm, dashed]
    ($(pushdown)+(-1.6,1)$) rectangle ($(fusion)+(2,-1.4)$);
    \node[color=agg_color, anchor=west, scale=1.1] at ($(pushdown)+(-1.2,-1.1)$)
    {\bf  Aggregate Optimisations};

    \node[data,draw=trie_color,color=trie_color] at ($(aggs) + (1,-6.5)$) (trie) {Dictionary\\ Nesting};
    \node[data,draw=trie_color,color=trie_color] at ($(trie) + (3.8,0)$) (fact2) {Factorisation/\\Code Motion};

    \node[data,draw=dl_color,color=dl_color] at ($(fact2) + (4.8,0)$) (dl1) {Physical\\Data-Structure};

    \node[data,draw=dl_color,color=dl_color] at ($(dl1) + (3.8,0)$) (layout) {Storage \\ Layout};
    \node[prog] at ($(layout) + (3,0)$) (cpp) {Target \\ Code};

    \draw[draw=trie_color, rounded corners = .1cm, dashed]
    ($(trie)+(-1.7,1)$) rectangle ($(fact2)+(2,-1.4)$);
    \node[color=trie_color, anchor=west, scale=1.1] at ($(trie)+(-1.2,-1.1)$)
    {\bf  Trie Conversion};

    \draw[draw=dl_color, rounded corners = .1cm, dashed]
    ($(dl1)+(-2.1,1)$) rectangle ($(layout)+(1.5,-1.4)$);
    \node[color=dl_color, anchor=west, scale=1.1] at ($(dl1)+(-1.8,-1.1)$)
    {\bf  Data Layout Synthesis};
    

    \draw[path] (aggs) -- (sched);
    \draw[path] (sched) -- (factor);
    \draw[path] (factor) -- (memo);
    \draw[path] (memo) -- (motion);

    \draw[path] (motion)  -- ++(0,-1.7) --  ++(-14,0) -- (unroll);

    \draw[path] (unroll) -- (static);
    \draw[path] (static) -- (pushdown);
    \draw[path] (pushdown) -- (fusion);

    \draw[path] (fusion) -- ++(3, 0) -- ++(0,-1.7) --  ++(-14.55,0) -- (trie);

    \draw[path] (trie) -- (fact2);
    \draw[path] (fact2) -- (dl1);
    \draw[path] (dl1) -- (layout);
    \draw[path] (layout) -- (cpp);
    
  \end{tikzpicture}
\vspace{3.5cm}

\caption{Transformation steps in IFAQ.}	
\label{fig:ifaq}
\end{figure}
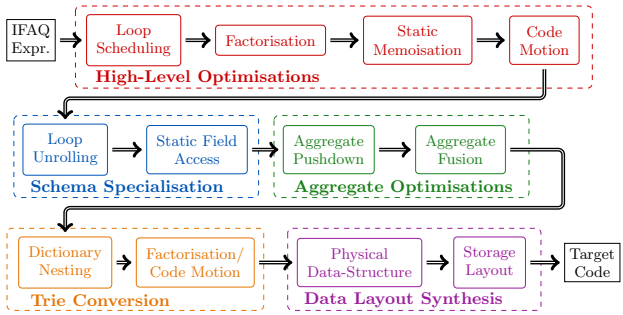

\begin{quote}
	{\em IFAQ can automatically synthesise and optimise aggregates from ML+DB workloads.} 
\end{quote}

We exemplify IFAQ~\cite{IFAQ:CGO:2020} using a simplified gradient descent program that learns the parameters of a linear regression model over the join $\col{Q}=S\bowtie R \bowtie I$ of relations $S(i, s, u)$, $R(s, c)$, and $I(i, p)$\footnote{The relations can be for instance \underline{S}ales(\underline{i}tem, \underline{s}tore, \underline{u}nits), Sto\underline{R}es(\underline{s}tore, \underline{c}ity), and \underline{I}tems(\underline{i}tem, \underline{p}rice).}. The model predicts the response $u$ given the set of numerical features $\col{F} = \{i, s, c, p\}$.

IFAQ takes the gradient descent program through the sequence of equivalence-preserving rule-based transformations shown in Figure~\ref{fig:ifaq}. We start with the program:

\begin{small}
\begin{align*}
  &\letbinding{\col{F}}{\{i, s, p, c\}}{}\\
  &\colm{\theta} \leftarrow \colm{\theta_0}\\
  &\code{while(}\whilecond{}\code{)\;\{}\\
  &\hspace*{.25em}
    \colm{\theta} = \dictbuild{f_1 \in \col{F}}{}\Bigg(\colm{\theta}(f_1) -
    \summation{\colm{x} \in \dom{\col{Q}}}{} \hspace{-0.5em}
    \col{Q}(\colm{x}) * \Big(\summation{f_2 \in \col{F}}{}
    \colm{\theta}(f_2)  * \colm{x}(f_2) \Big)* \colm{x}(f_1)\Bigg)\\[-0.2em]
  &\code{\}}\\
  &\colm{\theta}
\end{align*}
\end{small}
IFAQ supports collections via dictionaries, which map keys to values that can be simple data types or again dictionaries. Sets are dictionaries, where we only show the keys. Example of dictionaries are: $\dom{\col{Q}}$ mapping of tuples in the join $\col{Q}$ to feature vectors; the feature set $x$ mapping features ($i$, $s$, $p$, and $c$) to values; the corresponding parameter set $\colm{\theta}$. IFAQ has three iteration constructs: $\dictbuild{e \in set}{f(e)}$ constructs a dictionary that maps each key $e$ to $f(e)$; the stateful computation $\summation{e \in set}{f(e)}$ sums $f(e)$ over the elements in the set; and the $\code{while}$ construct is used for general iterations.

Without loss of generality, we left out the convergence condition and the step size and assume that the response $u$ is also included in $\colm{x}$ ($\colm{\theta}(u) = 1$ need not be updated).

\medskip

{\noindent\bf High-level optimisations.} We first normalise the gradient by bringing $\col{Q}(x)$ and $\col{x}(f_1)$ inside the innermost summation, swap the two summations so that the inner one iterates over larger dictionaries than the outer one, and factor out $\colm{\theta}(f_2)$ to use less arithmetic operations. The update step becomes:
\begin{small}
\begin{align*}
    \colm{\theta} = \dictbuild{f_1 \in \col{F}}{}\Bigg(\colm{\theta}(f_1) -
    \summation{f_2 \in \col{F}}{}
    \colm{\theta}(f_2) *\Big(\summation{\col{x} \in \dom{\col{Q}}}{}
     \col{Q}(x) * \col{x}(f_2) * \col{x}(f_1)\Big)\Bigg)
\end{align*}
\end{small}
Since the data-intensive computation over $\col{Q}$ in the above expression is independent of the two loops over $\col{F}$, we can hoist it outside the convergence loop. To achieve this, we first use memoisation to synthesise the dictionary $\col{M}$:
\begin{small}
\begin{align*}
 &\letbinding{\col{M}}{\dictbuild{f_1 \in \col{F}}{}\ \dictbuild{f_2 \in \col{F}}{}\ \summation{\col{x} \in \dom{\col{Q}}}{}\col{Q}(\col{x}) * \col{x}(f_2) * \col{x}(f_1)}{}\\
 &\colm{\theta} = \dictbuild{f_1 \in \col{F}}{}
       \Big(\colm{\theta}(f_1) - \summation{f_2 \in \col{F}}{}
       \colm{\theta}(f_2) * \col{M}(f_1)(f_2)\Big)
\end{align*}
\end{small}
The dictionary $\col{M}$ encodes the (non-centred) covariance matrix. Next, the code motion transformation moves the computation of $\col{M}$ outside the $\code{while}$ convergence loop:
\begin{small}
\begin{align*}
  &\letbinding{\col{F}}{\{i, s, p, c\}}{}\\
  &\letbinding{\col{M}}{\dictbuild{f_1 \in \col{F}}{}\ \dictbuild{f_2 \in \col{F}}{}\ \summation{\col{x} \in \dom{\col{Q}}}{}\col{Q}(\col{x}) * \col{x}(f_2) * \col{x}(f_1)}{}\\
  &\colm{\theta} \leftarrow \colm{\theta_0}\\
  &\code{while(}\whilecond{}\code{)\;\{}\\
  &\quad
    \colm{\theta} = \dictbuild{f_1 \in \col{F}}{}
       \Big(\colm{\theta}(f_1) - \summation{f_2 \in \col{F}}{}
       \colm{\theta}(f_2) * \col{M}(f_1)(f_2)\Big)\\[-0.2em]
  &\code{\}}\\
  &\colm{\theta}
\end{align*}
\end{small}
{\noindent\bf Schema specialisation.} Since the set $\col{F}$ of features is known at compile time, we can unroll the loops over $\col{F}$ in the definitions for $\col{M}$ and $\colm{\theta}$. We show below the expanded definition of $\col{M}$:
\begin{small}
\begin{align*}
\col{M} = \Big\{\ c \rightarrow \big\{\ldots, p \rightarrow
          \summation{\col{x} \in \dom{\col{Q}}}{} \col{Q}(\col{x}) * \col{x}(c) * \col{x}(p), \ldots
          \big\}, \ldots \Big\}
\end{align*}
\end{small}
We next enforce static field access wherever possible. This means that we convert dictionaries over $\col{F}$ into records so that the dynamic accesses to the dictionary now become static accesses. For instance, the dictionary $\col{x}$ becomes the record $x$ and then the dynamic access $\col{x}(c)$ becomes the static access $x.c$. The program becomes:
\begin{small}
\begin{align*}
& \code{let } M = \Big\{ c = \big\{\ldots, p =
          \summation{x \in \dom{\col{Q}}}{} \col{Q}(x) * x.c * x.p, \ldots
          \big\}, \ldots \Big\}\\[-0.2em]
& \code{in} \\
  &\colm{\theta} \leftarrow \colm{\theta_0}\\
  &\code{while(}\whilecond{}\code{)\;\{}\\
  &\quad
    \theta = \Big\{ c = \theta.c -
          \Big( \ldots + \theta.c * M.c.c +
          \theta.p * M.c.p \ldots \Big), \ldots
           \Big\}\\[-0.2em]
  &\code{\}}\\
  &\theta
\end{align*}
\end{small}
{\noindent\bf Aggregate optimisations.} The next set of transformations focus on database optimisations. In particular, we extract the aggregates from $M$, push them past the joins in $\col{Q}$, and share them across several entries in $M$. IFAQ identifies all aggregates in $M$ and pulls them out of $M$: 
\begin{small}
\begin{align*}
          &{\color{oxfordblue}
            \letbinding{M_{cc}}{\summation{x \in \dom{\col{Q}}}{}
              \col{Q}(x) * x.c * x.c}{}
          }\\
          &{\color{dgreen}
            \letbinding{M_{cp}}{\summation{x \in \dom{\col{Q}}}{}
              \col{Q}(x) * x.c * x.p}{}
          }\\
          &\letbinding{M}{\big\{ \ c = \big\{\ldots, 
            \ c = {\color{oxfordblue}M_{cc}}, 
            \ p = {\color{dgreen}M_{cp}}, 
            \ldots \big\}, \ldots \big\}}{} \ldots
\end{align*}
\end{small}
We next unfold the definition of the join $\col{Q}$. Recall that we do not want to materialise this join for runtime performance reasons as argued in Section~\ref{sec:introduction}. IFAQ  represents relations as dictionaries mapping tuples to their multiplicities. The join query $\col{Q}$ can be expressed in IFAQ as follows:
\begin{small}
    \begin{align*}
	\col{Q} = &\summation{x_s \in \dom{\col{S}}}{} \ 
  			\summation{x_r \in \dom{\col{R}}}{} \ 
  			\summation{x_i \in \dom{\col{I}}}{} \Big(\\
  &\letbinding{k}{\{i=x_s.i,s=x_s.s,c=x_r.c,p=x_i.p\}}\\
  \{k \rightarrow &\col{S}(x_s)*\col{R}(x_r)*\col{I}(x_i)*
  (x_s.i==x_i.i)*(x_s.s==x_r.s)\}\Big) 
    \end{align*}
\end{small}    
By inlining $\col{Q}$ in {\color{dgreen}$M_{cp}$}, we obtain:
\begin{small}
    \begin{align*}
{\color{dgreen} M_{cp}} = &\summation{x_s \in \dom{\col{S}}}{} \ 
  							\summation{x_r \in \dom{\col{R}}}{} \ 
  							\summation{x_i \in \dom{\col{I}}}{} \Big(\\
  &\letbinding{k}{\{i=x_s.i,s=x_s.s,c=x_r.c,p=x_i.p\}}\\
  &\code{let } \col{Q} = 
	\{k \rightarrow \col{S}(x_s)*\col{R}(x_r)*\col{I}(x_i)* \\
  & \hspace*{6.5em} (x_s.i==x_i.i)*(x_s.s==x_r.s)\} \code{ in } \\
  &{\color{dgreen} \summation{x\in\dom{\col{Q}}} {\col{Q}(x) * x.c * x.p}}\Big)
    \end{align*}
\end{small}    
Since $\dom{\col{Q}}=\{k\}$ is a singleton set, we can replace the last line above by $\col{Q}(k) * x.c * x.p$. Further inlining field accesses and retrieving the value of a singleton dictionary yields the following expression:
\begin{small}
    \begin{align*}
{\color{dgreen} M_{cp}} = &\summation{x_s \in \dom{\col{S}}}{} \ 
  							\summation{x_r \in \dom{\col{R}}}{} \ 
  							\summation{x_i \in \dom{\col{I}}}{} \Big(\\
  &\col{S}(x_s)*\col{R}(x_r)*\col{I}(x_i)* (x_s.i==x_i.i)*(x_s.s==x_r.s)  \\
  &{\color{dgreen} *\  x.c * x.p}\Big)
    \end{align*}
\end{small}    
We can now leverage the distributivity of multiplication over addition to factorise the expression:
\begin{small}
    \begin{align*}
{\color{dgreen} M_{cp}} = 
	&\summation{x_s \in \dom{\col{S}}}{} \col{S}(x_s) \ * \\ 
  	&\hspace*{1em} \summation{x_r \in \dom{\col{R}}}{} \col{R}(x_r) \ *\  (x_s.s==x_r.s) {\color{dgreen}\ *\  x_r.c} \ * \\ 
  	&\hspace*{2em} \summation{x_i \in \dom{\col{I}}}{} \col{I}(x_i) \ *\  (x_s.i==x_i.i){\color{dgreen} \ * \  x_i.p}
    \end{align*}
\end{small}
By static memoisation and loop-invariant code motion, we partially push the aggregates past the joins. We effect this using the dictionaries  $\col{V}_R$ and $\col{V}_I$ for the partial aggregates:
{\color{dgreen}
\begin{small}
    \begin{align*}
    \code{let } \col{V}_R &= \summation{x_r \in \dom{\col{R}}}{} \col{R}(x_r) \ *\ \{ \{s=x_r.s\} \rightarrow  x_r.c \} \code{ in}\\
    \code{let } \col{V}_I &= \summation{x_i \in \dom{\col{I}}}{} \col{I}(x_i) \ *\ \{ \{i=x_i.i\} \rightarrow  x_i.p \} \code{ in}\\
    \code{let } M_{cp} &= \summation{x_s \in \dom{\col{S}}}{} \col{S}(x_s) \ *\ \col{V}_R(\{s=x_s\}) \ * \ \col{V}_I(\{i = x_s.i\})
    \end{align*}
\end{small}
}
We can obtain {\color{oxfordblue}$M_{cc}$} similarly:
{\color{oxfordblue}
\begin{small}
    \begin{align*}
    \code{let } \col{V}_R' &= \summation{x_r \in \dom{\col{R}}}{} \col{R}(x_r) \ *\ \{ \{s=x_r.s\} \rightarrow  x_r.c \ * \ x_r.c \} \code{ in}\\
    \code{let } \col{V}_I' &= \summation{x_i \in \dom{\col{I}}}{} \col{I}(x_i) \ *\ \{ \{i=x_i.i\} \rightarrow  1 \} \code{ in}\\
    \code{let } M_{cc} &= \summation{x_s \in \dom{\col{S}}}{} \col{S}(x_s) \ *\ \col{V}_R'(\{s=x_s\}) \ * \ \col{V}_I'(\{i = x_s.i\})
    \end{align*}
\end{small}
}
The expressions for {\color{dgreen}$M_{cp}$} and {\color{oxfordblue}$M_{cc}$} are similar. To benefit from shared computation, we fuse them:
\begin{small}
    \begin{align*}
    \code{let } \col{W}_R &= \summation{x_r \in \dom{\col{R}}}{} \col{R}(x_r) \ *\ \{ \{s=x_r.s\} \rightarrow  \\
    &\hspace*{4em} \{{\color{dgreen}v_r = x_r.c}, {\color{oxfordblue}v_R' = x_r.c \ * \ x_r.c} \}\} \code{ in}\\
    \code{let } \col{W}_I &= \summation{x_i \in \dom{\col{I}}}{} \col{I}(x_i) \ *\ \{ \{i=x_i.i\} \rightarrow \\
    &\hspace*{4em} \{ {\color{dgreen}v_I = x_i.p}, {\color{oxfordblue}v_I' = 1} \} \} \code{ in}\\
    \code{let } M_{cc,cp} &= \summation{x_s \in \dom{\col{S}}}{} \col{S}(x_s) \ *\ \Big(\\
    \code{let } w_R &= \col{W}_R (\{s=x_s.s\}), w_I = \col{W}_I (\{i=x_s.i\}) \code{ in }\\
    &\hspace*{1em}\{{\color{dgreen}m_{cp} = w_R.v_R * w_I.v_I}, {\color{oxfordblue}m_{cc} = w_R.v_R' * w_I.v_I'} \}\Big)
    \end{align*}
\end{small}
Then {\color{oxfordblue}$M_{cc} = M_{cc,cp}.m_{cc}$} and {\color{dgreen}$M_{cp} = M_{cc,cp}.m_{cp}$}.

{\noindent\bf Trie conversion.} By representing $\col{S}$ as a nested dictionary, first grouped by $x_s$ and then by $x_i$, IFAQ factors out the multiplication of the elements in $w_R$ with those in $w_i$ before the enumeration of the $x_i$ values. This further reduces the number of arithmetic operations needed to compute $M_{cc,cp}$.

{\noindent\bf Data layout.} As data structures for dictionaries, IFAQ supports hash tables, balanced-trees, and sorted dictionaries. Each of them show advantages for different workloads.

\nop{

 Data-Layout Synthesis
 
 Turn resulting IFAQ expression into efficient low-level C++ code:
1. Represent static records as C++ structs
2. Turn immutable data-structures into mutual data-structures 3. Turn intermediate records into local variables
4. Replace single-field records by their single field
5. Turn dictionaries into arrays when possible

}

\section{Reflections}
\label{sec:reflections}

The work overviewed in this paper follows a three-step recipe for efficient learning over relational data:
\begin{enumerate}
\item Turn the learning problem into a database problem.
\item Exploit the problem structure to lower the complexity.
\item Generate optimised code to lower the constant factors.
\end{enumerate}

This recipe proved surprisingly effective for a good number of machine learning models and also reinvigorated the fundamental problems of query evaluation and optimisation. Further principled work on the theory and systems of structure-aware learning approaches is needed. What are the limits of structure-aware learning? What other classes of machine learning models can benefit from it? What other types of structure are relevant to the learning task?
A major target remains the development of data systems that can natively run database and machine learning workloads as one. A pragmatic step towards this goal is to integrate database techniques for feature extraction queries, such as those described in this paper, in popular tools used by data scientists, e.g., TensorFlow, scikit-learn, R, and jupyter notebooks. A second major target is to translate the rich knowledge readily available in curated relational databases into the accuracy of the trained models.

It may be a painstaking endeavour to understand and translate success stories on machine learning models and optimisation to the database coordinate system, though this may reveal deeper connections between the two fields and also draw them closer. Database research may also streng\-then the machine learning research landscape with its rigour in theoretical and system development, benchmarking, and reproducibility. Designing and teaching courses that present a more integrated and unified view of Computer Science fields, such as databases and machine learning, may equip both the lecturers and their students with the tools necessary to advance research and practice at the interface between seemingly disparate fields.

\balance

\bibliographystyle{abbrv}
\bibliography{main}


\end{document}